\documentclass{aa}  %

\usepackage{graphicx}
\usepackage{txfonts}
\usepackage{longtable}
\begin{document}

   \title{Understanding dust production and mass loss on the AGB phase using post-AGB stars in the Magellanic Clouds}


   \author{S. Tosi\inst{1,2}, F. Dell'Agli\inst{2}, D. Kamath$\inst{3,4}$, 
   P. Ventura\inst{2,5}, H. Van Winckel\inst{6}, E. Marini\inst{2}
          }

   \institute{Dipartimento di Matematica e Fisica, Università degli Studi Roma Tre, 
              via della Vasca Navale 84, 00100, Roma, Italy \and
              INAF, Observatory of Rome, Via Frascati 33, 00077 Monte Porzio Catone (RM), Italy \and
              School of Mathematical and Physical Sciences, Macquarie University, Sydney, NSW, Australia \and
              Astronomy, Astrophysics and Astrophotonics Research Centre, Macquarie University, Sydney, NSW, Australia \and
              Istituto Nazionale di Fisica Nucleare, section of Perugia, Via A. Pascoli snc, 06123 Perugia, Italy \and
              Institute of Astronomy, K.U.Leuven, Celestijnenlaan 200D bus 2401, B-3001 Leuven, Belgium
              }

   \date{Received September 15, 1996; accepted March 16, 1997}


 \abstract
   {The asymptotic giant branch (AGB) phase of evolution in low- and intermediate-mass stars is governed by poorly understood physical mechanisms, such as convection, mixing, dust production and mass loss, which play a crucial role 
   on determining the internal structure and the evolution of these stars. The spectra of post-asymptotic giant branch (post-AGB) stars hold critical chemical fingerprints that serve as exquisite tracers of the evolution, nucleosynthesis, and dust production during the AGB phase.}
   {We aim to understand the variation of the surface chemistry that occurs during the AGB phase by analysing results from observations of single post-AGB stars in the Magellanic Clouds. We also aim at reconstruct dust formation processes, that are active in the circumstellar envelope of AGB stars, occurring towards the end of the AGB phase and during the subsequent course of evolution when contraction to the post-AGB has begun.}
   {We study likely single post-AGB sources in the Magellanic Clouds that exhibit a double-peaked (shell-type) spectral energy distribution (SED). We interpret their SED by comparing with results from radiative transfer 
   calculations, to derive the luminosity and the dust content of the individual sources. Additionally, we compare the observationally derived stellar parameters and the photospheric chemical abundances of the target sample with results from stellar evolution modelling of AGB and post-AGB stars. This allows for the characterization of the individual sources in terms of initial mass and formation epoch of the progenitors. The theoretically derived dust mineralogy and optical depth is used to assess
   when dust formation ceases and determine the propagation velocity of the dust-gas system during the post-AGB evolution.}
   {We find that amongst our target sample of 13 likely single post-AGB stars with shell-type SED, 8 objects are carbon stars descending from $\sim 1-2.5~{\rm M}_{\odot}$ progenitors. 5 of the 13 objects are of lower mass, descending from ${\rm M} < 1~{\rm M}_{\odot}$ stars. Based on the dust mineralogy, we find that these 5 stars are surrounded by silicate dust, and thus failed to become carbon stars. The dust optical depth and the luminosity of the stars are correlated, owing
   to the faster evolutionary time-scale brighter stars, which makes the dusty layer to be closer to the central object. From our detailed analysis of the SEDs, we deduce that
   the dust currently observed around post-AGB stars was released after the onset of the central star contraction and an increase in the effective temperature to $\sim 3500-4000$ K.}
   {}


   \titlerunning{Evolution and dust formation in LMC carbon stars}
   \authorrunning{Tosi et al.}
   \maketitle
%

\section{Introduction}
After the core helium burning phase, single stars of initial mass below $\sim 8~{\rm M_{\odot}}$, i.e., 
low- and intermediate-mass stars, 
evolve through the asymptotic giant branch (AGB). When the mass of the envelope drops 
below a few hundredths of solar mass the external regions of the star contract, thus giving the 
start to the post-AGB evolution \citep{iben83a}.
During this phase the stars evolve at constant luminosity and readjust on a configuration
which becomes more and more compact as the time elapses, with the effective temperature increasing from
$\sim 3000$ K to $\sim 30000$ K. The post-AGB phase is then followed by the PN evolution, 
before the start of the white dwarf cooling \citep{iben83a}.

The AGB phase is characterized by the occurrence of two physical phenomena able to alter the surface chemistry of these stars: third dredge-up (TDU, Iben 1974) and hot bottom burning (HBB, Sackmann \& Boothroyd 1992). These mechanisms are still not fully understood from first principles and the information derived from the analysis and the characterization of post-AGB stars proves
an exceptionally valuable tool for this purpose. The main reason is twofold: their chemical composition represents the final outcome of AGB evolution and associated internal enrichment processes. Another reason is that the surface abundances of post-AGB stars can be determined accurately given that their spectra are dominated by atomic transitions.

Furthermore, given the peculiar morphology of the SED of post-AGB stars \citep{woods11}, 
the study of the infrared (IR) excess allows a reliable determination of the dust 
properties, in terms of mineralogy, optical thickness and current distance from the
central star: this kind of analysis provides important
information on the dust formation process in the wind of AGBs during the final evolutionary
phases, and more generally on the efficiency of dust formation in the 
circumstellar envelope of evolved giants. This step is of paramount importance to
evaluate the role that AGB stars play as dust manufacturers: indeed it was shown that
a few stars evolving through the late AGB phases are responsible for most of the current 
dust formation rate in the dwarf irregular galaxies like IC1613 \citep{flavia16}, 
IC10 \citep{flavia18}, Sextans A \citep{flavia19}. 

In the present work we use results from observations of post-AGB stars to characterize the
individual sources in terms of mass and age of the progenitor stars. This step offers the
possibility of testing the current AGB evolution theories, particularly when the surface
chemistry of the stars is known from the observations, and can be compared with the
theoretical predictions. Furthermore, the analysis of the IR excess is used to
shed new light on the dust formation process during the very late AGB phases and possibly
after the beginning of the contraction to the post-AGB. This investigation can draw information on the
time-scale of the AGB to post-AGB transition, and to investigate the dynamics of the wind 
moving away from the star after dust production ceases.

This study benefits from the rich dataset of optically visible post-AGB stars
observed in the Magellanic Clouds (Kamath et al. 2014, 2015, hereafter K14 and K15).
For the single stars in these samples the availability of the optical and infrared data allowed the reconstruction of the spectral energy distribution (SED) and the determination of the IR excess, 
which prove essential for an estimate of the luminosity and for the determination of the mineralogy and
amount of dust in the surroundings of the star.

The stars are interpreted and characterized under the light of updated AGB evolutionary
sequences, extended on purpose until and past the post-AGB phase. These tracks also include the
description of the dust formation process, following the schematization proposed by the
Heidelberg group \citep{fg06}, on the wake of previous studies on dust production by evolved stars
proposed by our team \citep{ventura12, ventura14}, which have been used
to interpret the evolved stellar populations of the Magellanic Clouds \citep{flavia14a,
flavia14b, flavia15a, flavia15b} and of Local Group galaxies \citep{flavia16, flavia18, flavia19},
and to explore the potentialities in this field offered by the recent launch of the
James Webb Space Telescope \citep{ester20, ester21}.

The paper is structured as follows: Section \ref{input} includes the description of the 
codes used to model the evolution of the star and the dust formation process during the
AGB and the early post-AGB phases, and the methodology followed to interpret the
optical and infrared observations of the selected sources, to derive the luminosities
and the properties of the dust in their surroundings; the results of the SED fitting analysis
are given in section \ref{sedfit}; in section \ref{disc} we characterize the sources
individually, in order to find out the mass and age of the progenitors; section \ref{lateagb}
presents results regarding the mass loss suffered by the stars during the late AGB phases,
in agreement with the expectations from stellar evolution modelling;
in section \ref{wind}  we discuss the properties of the wind and of the physics of the 
AGB - post-AGB transition, based on the results obtained in the previous sections. Finally, 
the conclusions are given in section \ref{concl}.

\section{Numerical, chemical and physical input}
\label{input}
The characterization of the stars considered in the present study is based on
results from stellar evolution and dust formation modelling, which are used
in synergy with SED fitting analysis, applied to all the sources investigated.
In the following we describe the most relevant physical ingredients of the codes
used to model the evolution of the star, to describe the dust formation process
in the outflow, and to build synthetic SEDs, which are compared with the
observational data set available.

\subsection{AGB and post-AGB modelling}
\label{agbmod}

The evolutionary sequences were calculated by means of
the stellar evolution code ATON, extensively described in \citet{ventura98}.
The code performs a full integration of the stellar structure equations, from the
centre of the star to the photosphere. The temperature gradient within regions unstable
to convection is found via the Full Spectrum of Turbulence model \citep{cm91}.
The formal borders of the convective regions are found via the classic Schwartzschild 
criterion; however, convective eddies are allowed to overshoot into radiatively
stable regions, by imposing an exponential decay of velocities beyond the formal border.
This approach, which is consistent with results from numerical simulations \citep{freytag},
was also used e.g. by \citet{herwig00}.
In the convective regions (including the overshoot zones) nuclear burning and mixing of
chemicals are self-consistently coupled by means of the diffusive approach proposed
by \citet{cloutman}. Mass loss during the oxygen-rich phases is described by using
the formulation by \citet{blocker95}; for carbon stars we use the treament proposed by
the Berlin group \citep{wachter02, wachter08}.

The metallicity of the stars considered here span the 
$-1 \leq [{\rm Fe}/{\rm H}] \leq -0.4$ range. We divided the sources in two 
groups, according to the metallicity, and used the $Z=2\times 10^{-3}$ and $Z=4\times 10^{-3}$ 
evolutionary tracks that we calculated for the study by Kamath et al. (2021),
to study the stars of lower and higher metallicity, 
respectively. To investigate the highest luminosity objects we extended 
the set of models by Kamath et al. (2021) to include $2~{\rm M}_{\odot}$ and $2.5~{\rm M}_{\odot}$
evolutionary sequences of metallicity $Z=2\times 10^{-3}$, which were calculated on purpose for
the present work. Furthermore, the study of the faintest
stars required the calculation of $0.65~{\rm M}_{\odot}$ and $0.7~{\rm M}_{\odot}$
models of the same metallicity.

\subsection{The description of dust formation in the wind of evolved stars}
\label{dustmod}
For some selected points along the evolutionary tracks we use the physical
parameters of the star, namely mass, luminosity, effective temperature, mass loss rate
and surface chemical composition, to determine the dust production rate. To this aim
we rely on the schematization proposed by the Heidelberg group \citep{fg06},
according to which the dust forms in a stationary wind, expanding radially from the
photosphere of the star. 

The wind velocity $v$ is found via the equation of momentum conservation, which is

$$
{d{\rm v}\over dr}=-{G{\rm M}_*\over r^2}(1-\Gamma)
\eqno(1)
$$

where $r$ is the radial distance from the center of the star, ${\rm M}_*$ is the (current) mass of the star and $\Gamma={k{\rm L}_*/4\pi cG{\rm M}_*}$. In the 
expression for $\Gamma$, which represents the relative weight of the effects of radiation
pressure and gravity, ${\rm L}_*$ is the luminosity of the star, whereas $k$ is the extinction
coefficient, which reflects the scattering and absorption processes of the photons
released from the stellar photosphere by dust particles.

Therefore, the growth of dust grains and the dynamics of the wind are self-consistently 
coupled, as the calculation of $k$ requires the determination of the size of the dust particles
of the different species formed.

The description of the wind is completed by the radial profiles of the density,
which for mass conservation reasons is expressed by the equation

$$
\rho={\dot {\rm M}\over {4\pi r^2 {\rm v}}}
\eqno(2)
$$

where $\dot {\rm M}$ is the mass loss rate, whereas v is the wind velocity at distance $r$ from the
centre of the star.

The relevant equations and the details of the numerical treatment are extensively described, 
e.g., in \citet{ventura12, ventura14}.

The knowledge of the radial stratification of the thermodynamic variables and
of the grain sizes of the various dust species allows the calculation of the optical depth, 
which is found via the following expression:

$$
\tau_{10}=\pi\int_{r_{in}}^{r_{out}}{n_dQ_{10}a^3dr}
\eqno{(3)}
$$

where $n_d$ and $a$ are the number density and the size of the dust grains, $Q_{10}$ is the
extinction coefficient at $10~\mu$m, $r_{in}$ and $r_{out}$ are the distances of the inner and outer 
borders of the region of the circumstellar envelope populated by dust particles. The 
overall optical depth must be calculated by summing all the contributions from the individual 
dust species, the largest contribution being given by those with the highest extinction coefficients,
i.e. solid carbon or silicate dust. 

\subsection{SED Fitting}
\label{sedmod}
To build synthetic SED for the comparison with the observational data points
we use the radiation transfer code DUSTY, described in \citet{nenkova99}.
The code calculates the SED emerging from the star, by accounting for the
reprocessing of the radiation released from the photosphere by a dusty region,
isotropically distributed around the source.

To describe the input radiation entering the dusty layer we use the
Kurucz-Castelli atmosphere models \citep{castelli}, available for a wide
range of metallicities and effective temperatures.

The effects of dust reprocessing are taken into account by specifying the
optical depth $\tau_{10}$ (see Eq.~3 above for the definition) and the
dust temperature in the inner border of the dusty region, ${\rm T}_{\rm d}$.

For what regards the dust mineralogy, we consider the possibilities that the dust is 
composed either by $100\%$ silicates
or it is pure carbon dust. In the latter case, for the higher metallicity sources,
we further explore the possibility that some fraction of silicon carbide is 
present. 

The size of the dust particles is chosen on the basis of the results from
dust formation modelling, described in section \ref{dustmod}. One of the outcomes
of this modelling is the asymptotic size attained by the dust grains of the
individual species, which is used as input for the dusty code. Typically, we adopt
$0.12~\mu$m for solid carbon, $0.07~\mu$m for SiC, and $0.08~\mu$m for silicates.

The optical constants required as input by DUSTY and considered in the present
analysis are the following: \citet{oss92} for silicates; \citet{zubko} for amorphous carbon;
\citet{peg} for silicon carbide; we will consider the possibility that in the low temperature
domain the silicates assume a crystalline form, in which case we use the optical constants by
\cite{jager94}.

\section{The analysis of SEDs of the target sample of post-AGBs}
\label{sedfit}

\begin{figure*}
\begin{minipage}{0.32\textwidth}
\resizebox{1.\hsize}{!}{\includegraphics{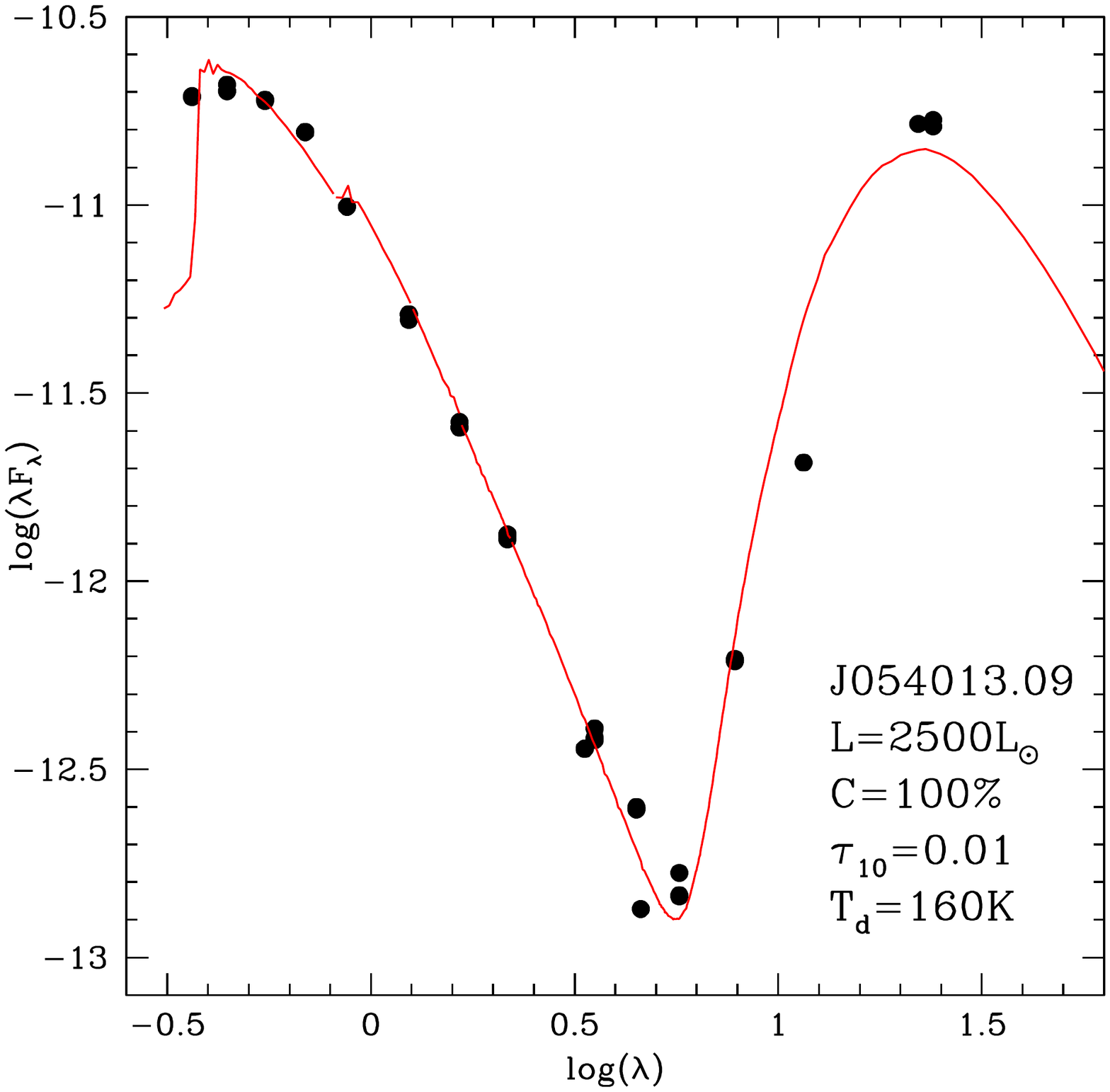}}
\end{minipage}
\begin{minipage}{0.32\textwidth}
\resizebox{1.\hsize}{!}{\includegraphics{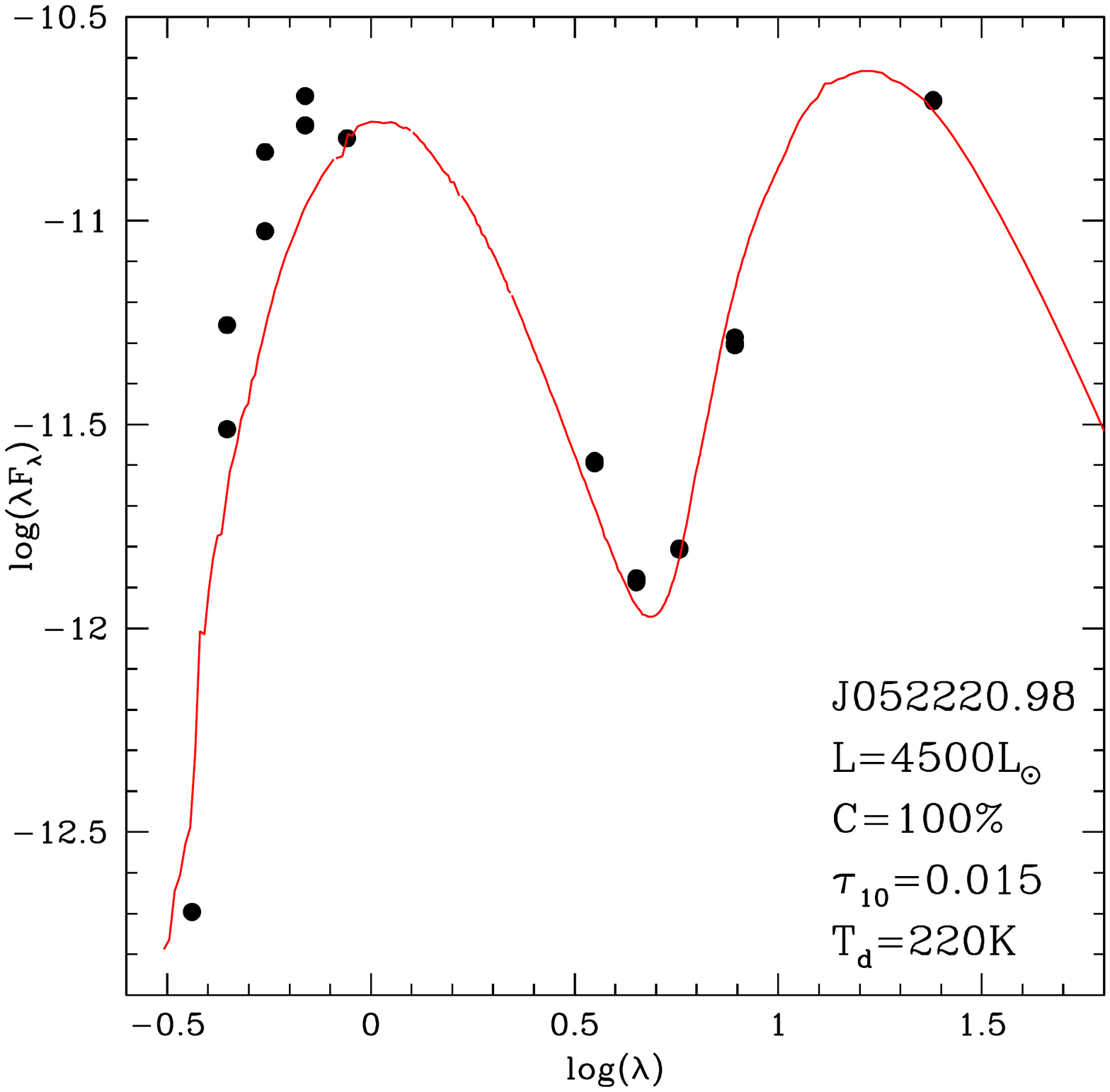}}
\end{minipage}
\begin{minipage}{0.32\textwidth}
\resizebox{1.\hsize}{!}{\includegraphics{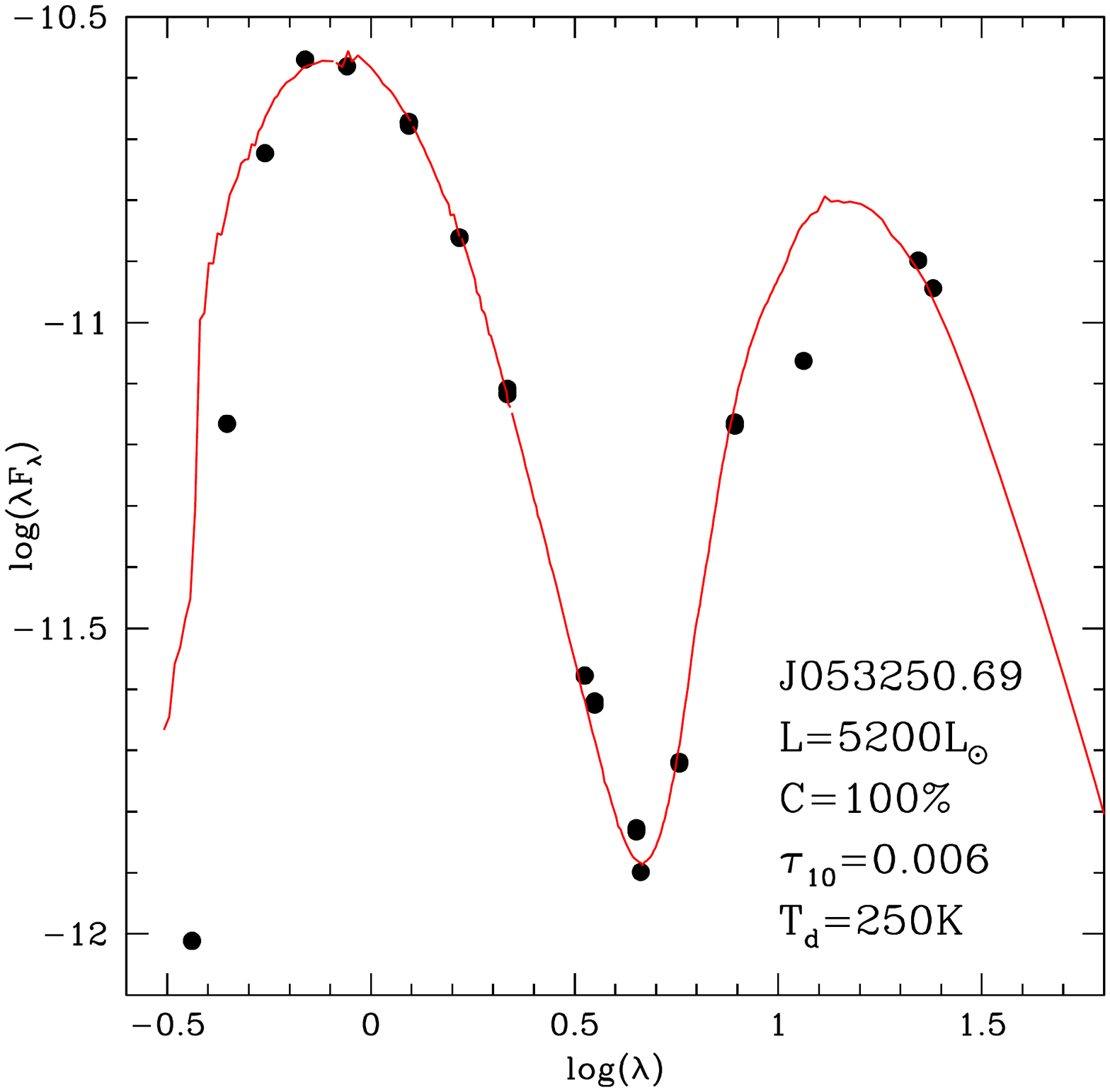}}
\end{minipage}
\vskip-70pt
\begin{minipage}{0.32\textwidth}
\resizebox{1.\hsize}{!}{\includegraphics{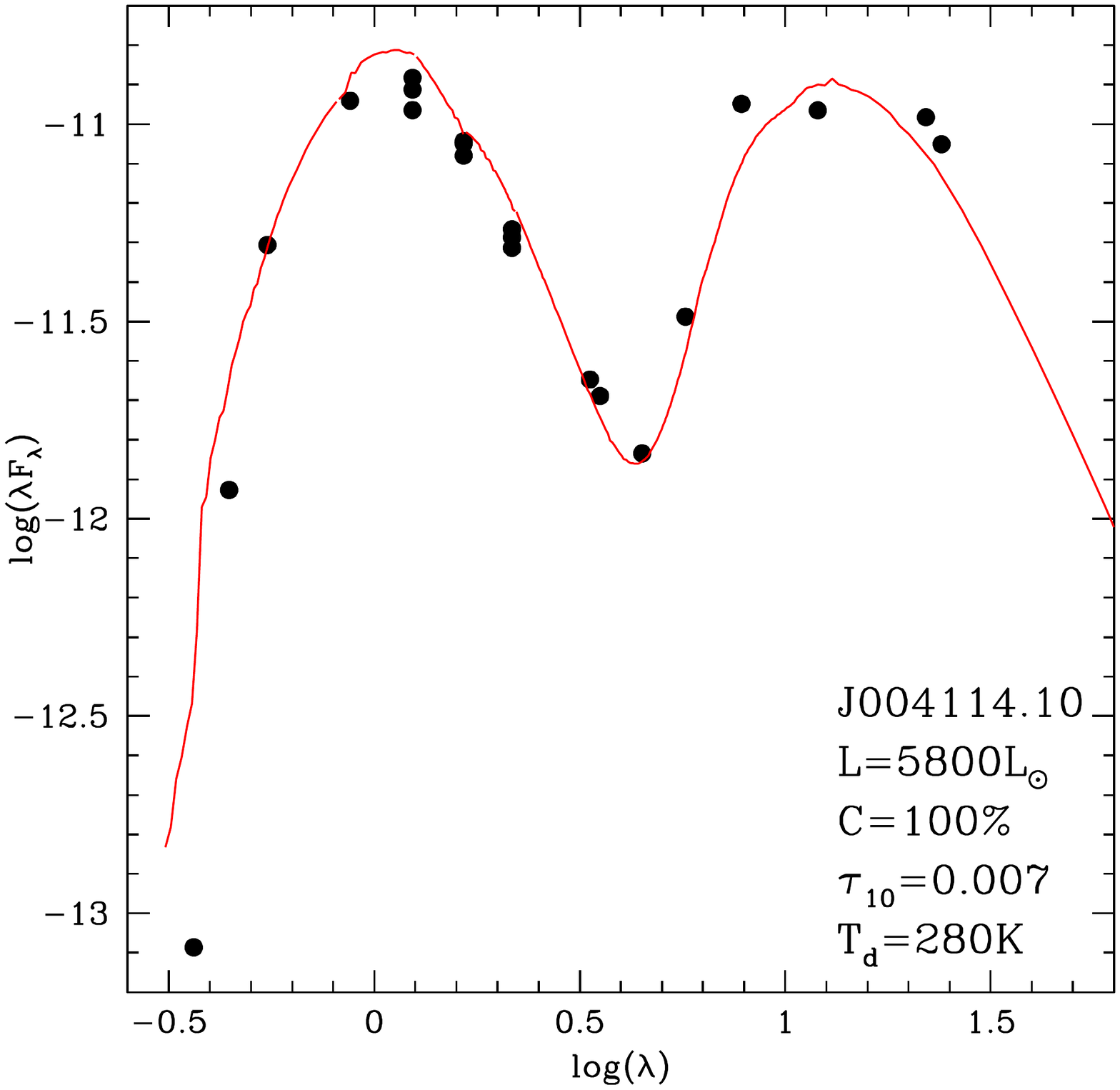}}
\end{minipage}
\begin{minipage}{0.32\textwidth}
\resizebox{1.\hsize}{!}{\includegraphics{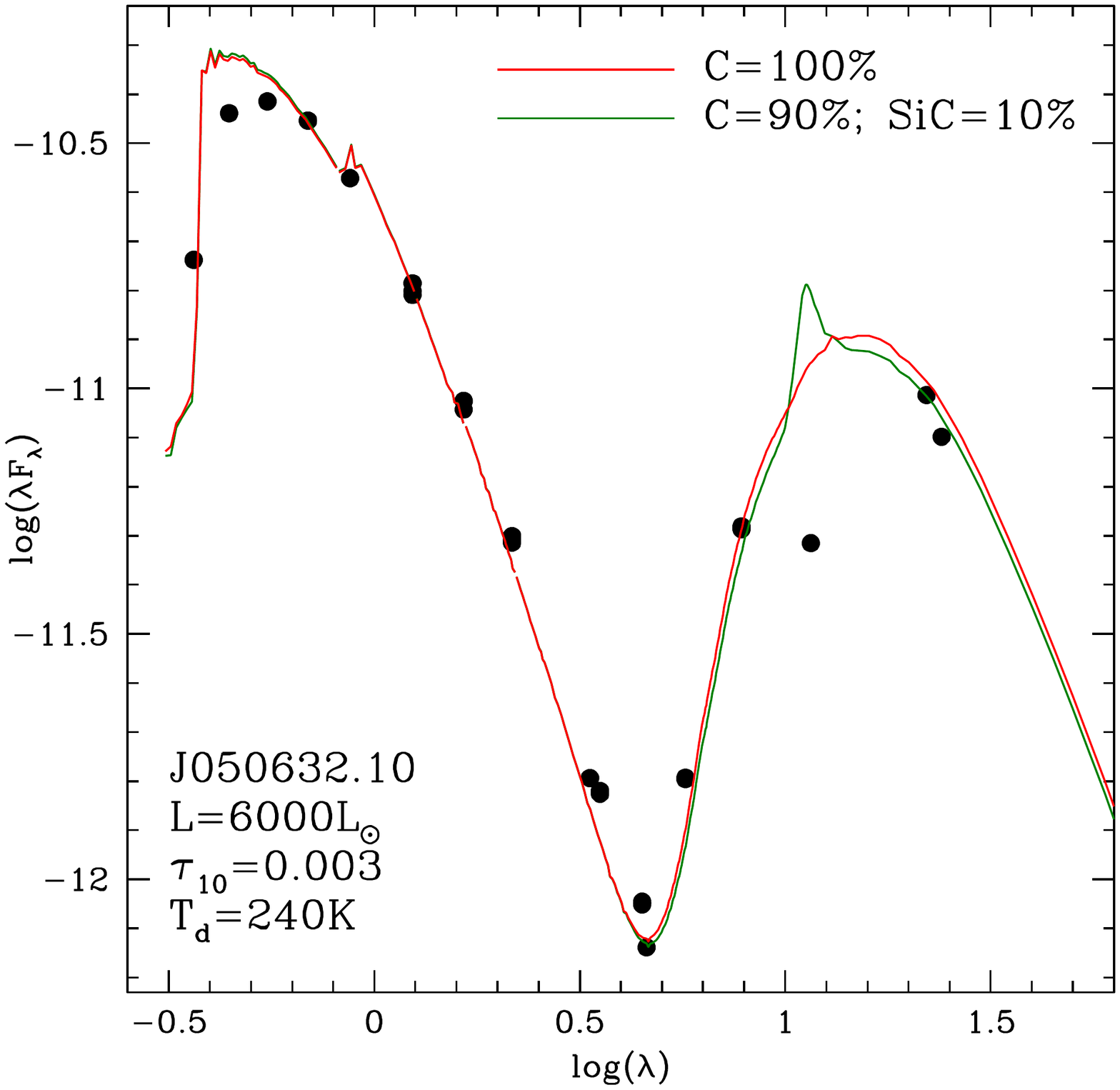}}
\end{minipage}
\begin{minipage}{0.32\textwidth}
\resizebox{1.\hsize}{!}{\includegraphics{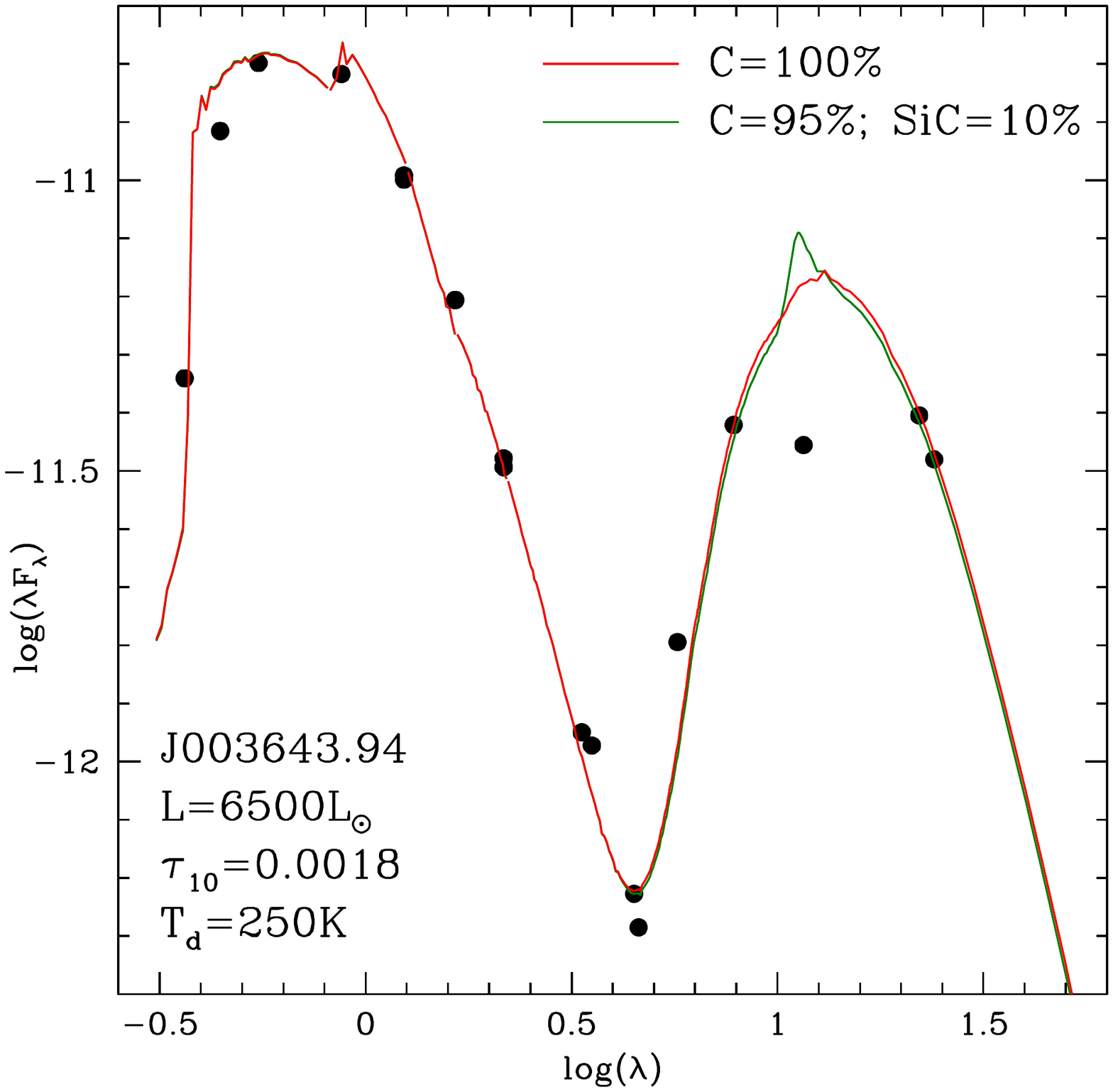}}
\end{minipage}
\vskip-70pt
\begin{minipage}{0.32\textwidth}
\resizebox{1.\hsize}{!}{\includegraphics{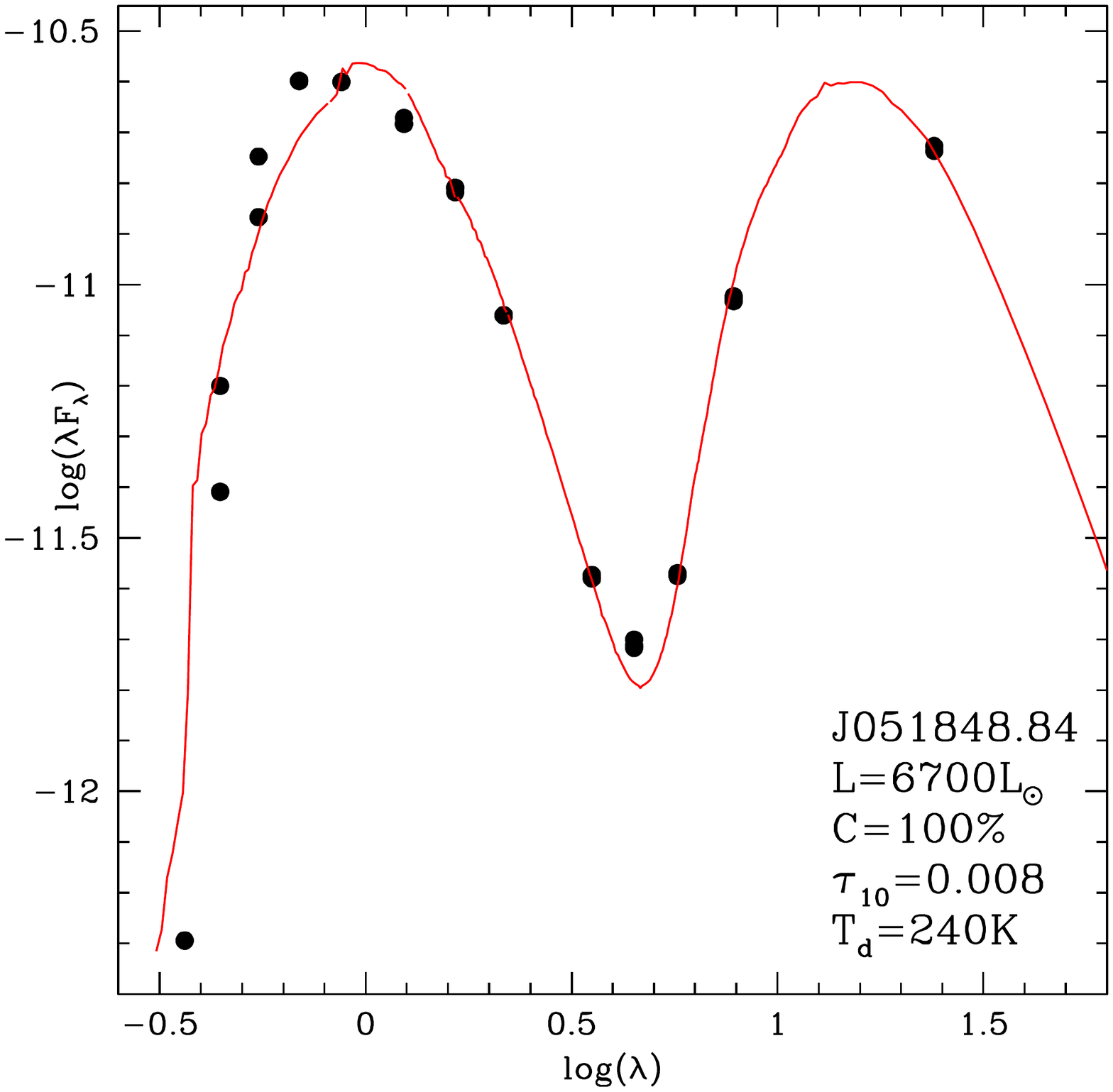}}
\end{minipage}
\begin{minipage}{0.32\textwidth}
\resizebox{1.\hsize}{!}{\includegraphics{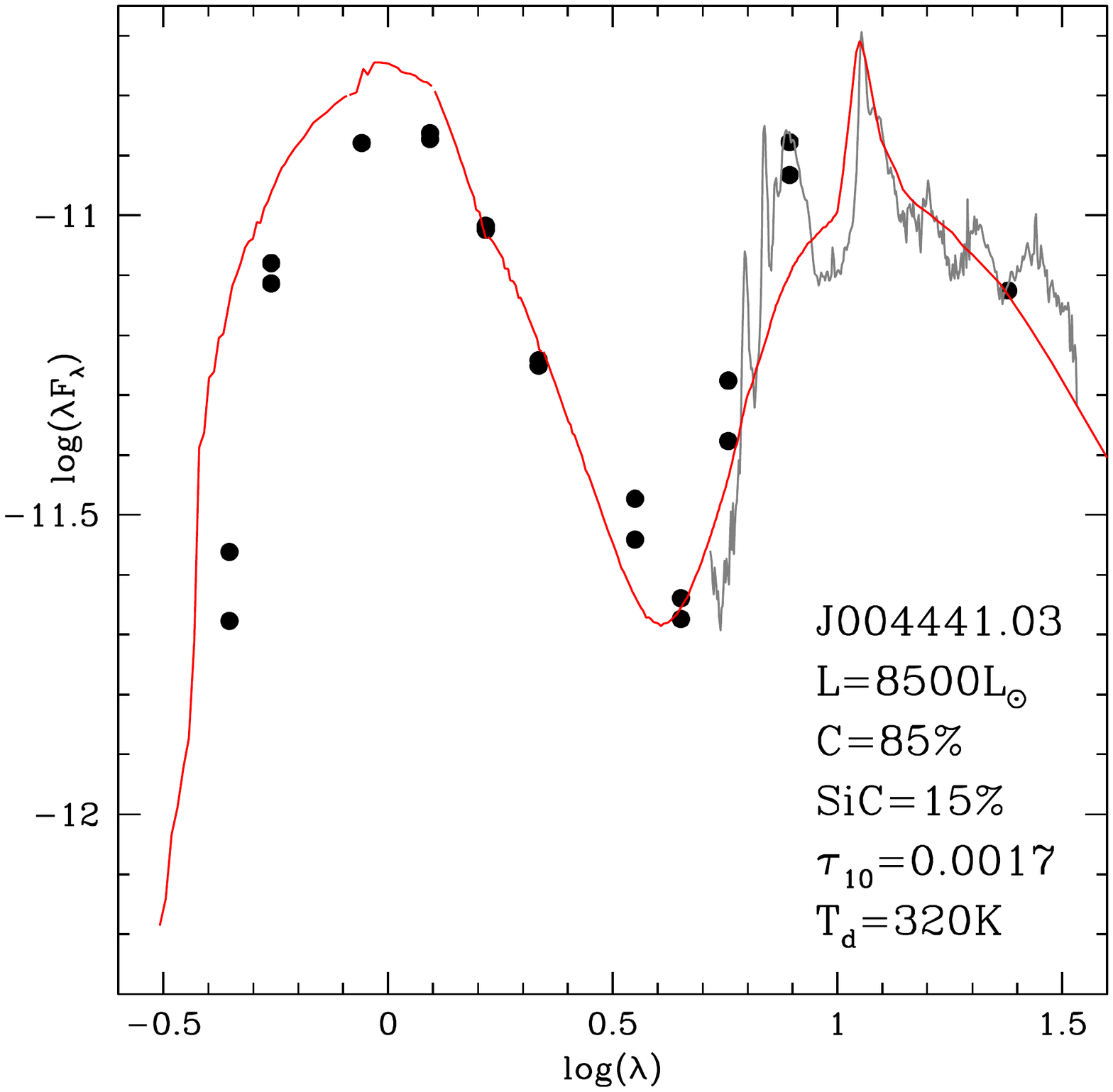}}
\end{minipage}
\begin{minipage}{0.32\textwidth}
\resizebox{1.\hsize}{!}{\includegraphics{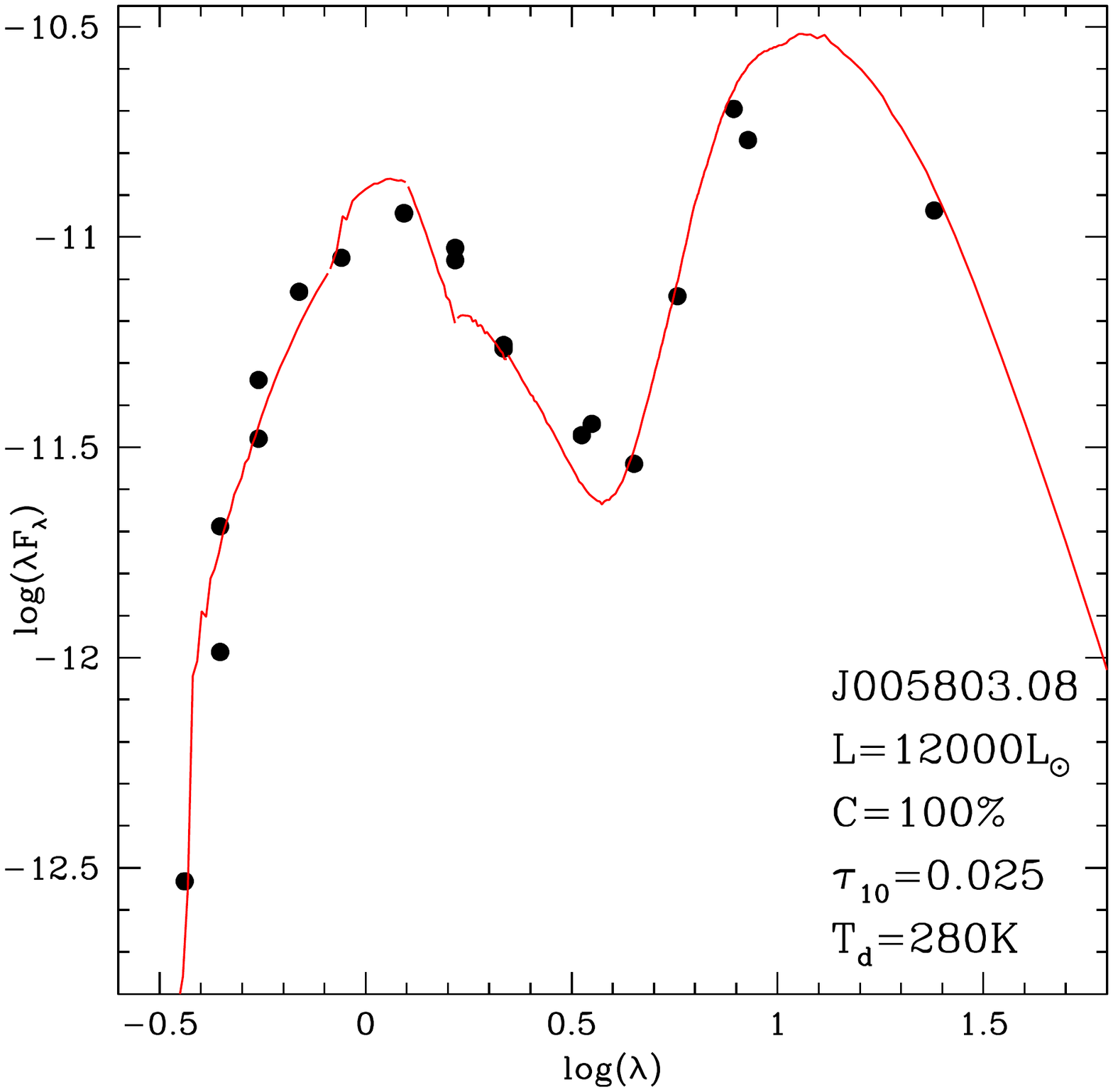}}
\end{minipage}
\vskip-40pt
\caption{Optical and IR data (black points) of SMC and LMC sources classified as post-AGBs with
a shell-like structure by K14 and K15, that we interpret as surrounded by carbonaceous dust in 
the present investigation. The grey line in the middle panel of the bottom line refer to
IR spectroscopy results from \citet{volk11}. The red lines indicate the best-fit model obtained by the
DUSTY code. The derived stellar and dust parameters for each source are shown in the
different panels.}
\label{fcstar}
\end{figure*}

K14 and K15 assembled results from optical and infrared photometry of evolved stars
in the Magellanic Clouds, and reconstructed the shape of the SED of the different sources over 
an extended spectral region. In this work we focus on 13 single stars that were classified as shell-type 
post-AGBs by K14 and K15. This choice allows us to use the code DUSTY, described in section
\ref{sedmod}, in the attempt of providing an interpretation of the data of the different
sources.

Our approach is to confront results from synthetic SED modelling with the photometric data set,
to derive the most relevant quantities related to the individual stars, namely luminosity,
dust mineralogy, optical depth and dust temperature of the internal border of the dusty region.
For only one of the sources considered, namely J004441.03, we could compare the synthetic SED with the IR
spectrum \citep{volk11}. 

To this aim we start from the effective temperatures and metallicities reported in K14 and K15, or,
when available, from high-resolution spectroscopy analysis. These information are required to
calculate the input spectrum released from the photosphere of the star, and entering the
dusty zone. We adjust the input parameters required to produce the synthetic SED, until satisfactory
agreement with the observations is reached. This procedure leads to the derivation of the
stellar luminosity and of the dust quantities, according to the following steps:

\begin{enumerate}

\item{The dust mineralogy can be reliably assessed, because
the shape of the SED of carbon stars differs significantly from that of stars surrounded by 
silicates in the $20-30 ~ \mu$m spectral region: indeed the presence of silicates
causes a strongly negative slope of the flux with $\lambda$ at wavelengths above $\sim 20~\mu$m, 
whereas in the case of carbon stars the SED is still decreasing in the same spectral region, 
but it is significantly flatter.}

\item{The optical depth of the dust surrounding the individual sources can be accurately derived, 
because the dust layer is sufficiently far from the surface of the
star, that the infrared excess in the $8-30 ~ \mu$m spectral region is indeed dependent 
on $\tau_{10}$ only.}

\item{The peculiar morphology of the SED of post-AGB stars allows the determination of the
dust temperature ${\rm T}_{\rm d}$ of the internal border of the dust zone, which turns extremely sensitive
to the depth and the location of the minimum in the SED, usually found in the $3-6~\mu$m
spectral region.}

\item{The morphology of the SED in the optical part of the spectrum is used to
derive the values of the interstellar reddening, for which we find consistency with
the results published in K14 and K15.}
    
\item{The luminosity of the star can be estimated by shifting the synthetic SED until
matching the observed SED profile in the near infrared region.}

\end{enumerate}

Once the mineralogy of the dust is determined, the dust temperature can be derived with 
an accuracy of the order of $10-20$K. In the cases where WISE, IRAC and MIPS photometry are 
available, the optical depth can be fixed with a $10-15\%$ precision. Finally, the luminosity 
of the stars is established mostly from the near-IR part of the spectrum with an uncertainty 
within $10\%$.

\begin{figure*}
\begin{minipage}{0.32\textwidth}
\resizebox{1.\hsize}{!}{\includegraphics{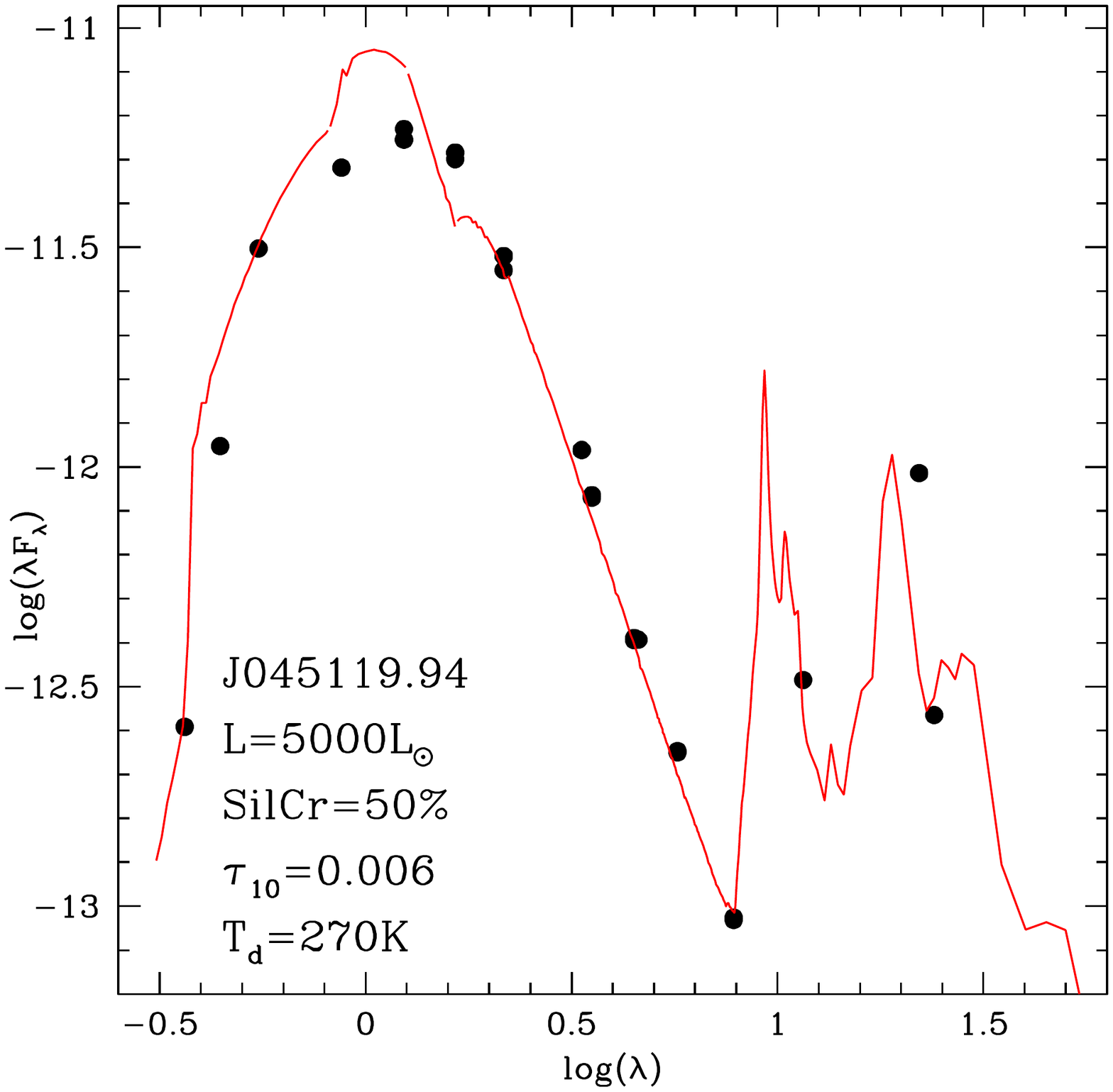}}
\end{minipage}
\begin{minipage}{0.32\textwidth}
\resizebox{1.\hsize}{!}{\includegraphics{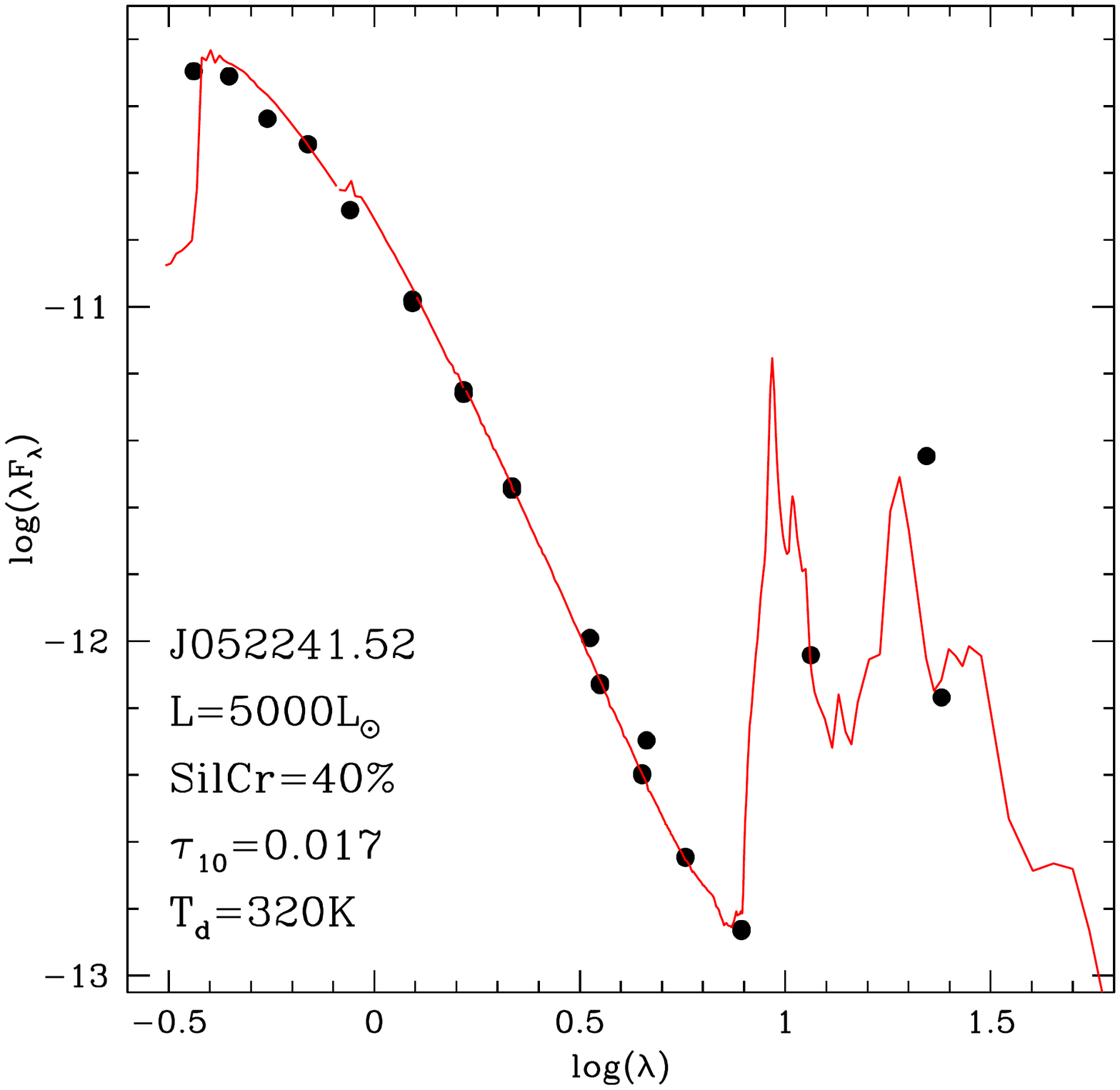}}
\end{minipage}
\begin{minipage}{0.32\textwidth}
\resizebox{1.\hsize}{!}{\includegraphics{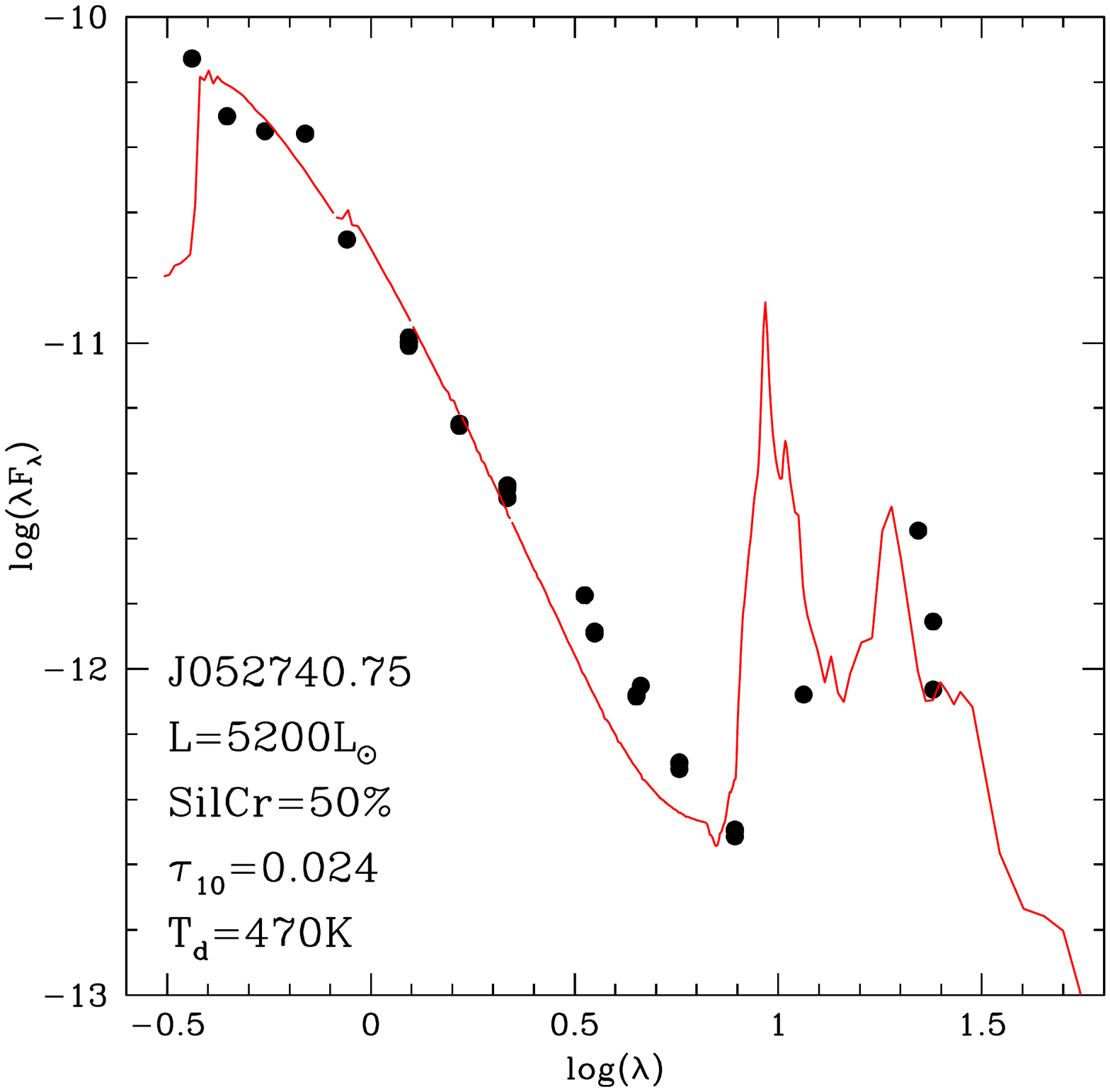}}
\end{minipage}
\vskip-70pt
\begin{minipage}{0.32\textwidth}
\resizebox{1.\hsize}{!}{\includegraphics{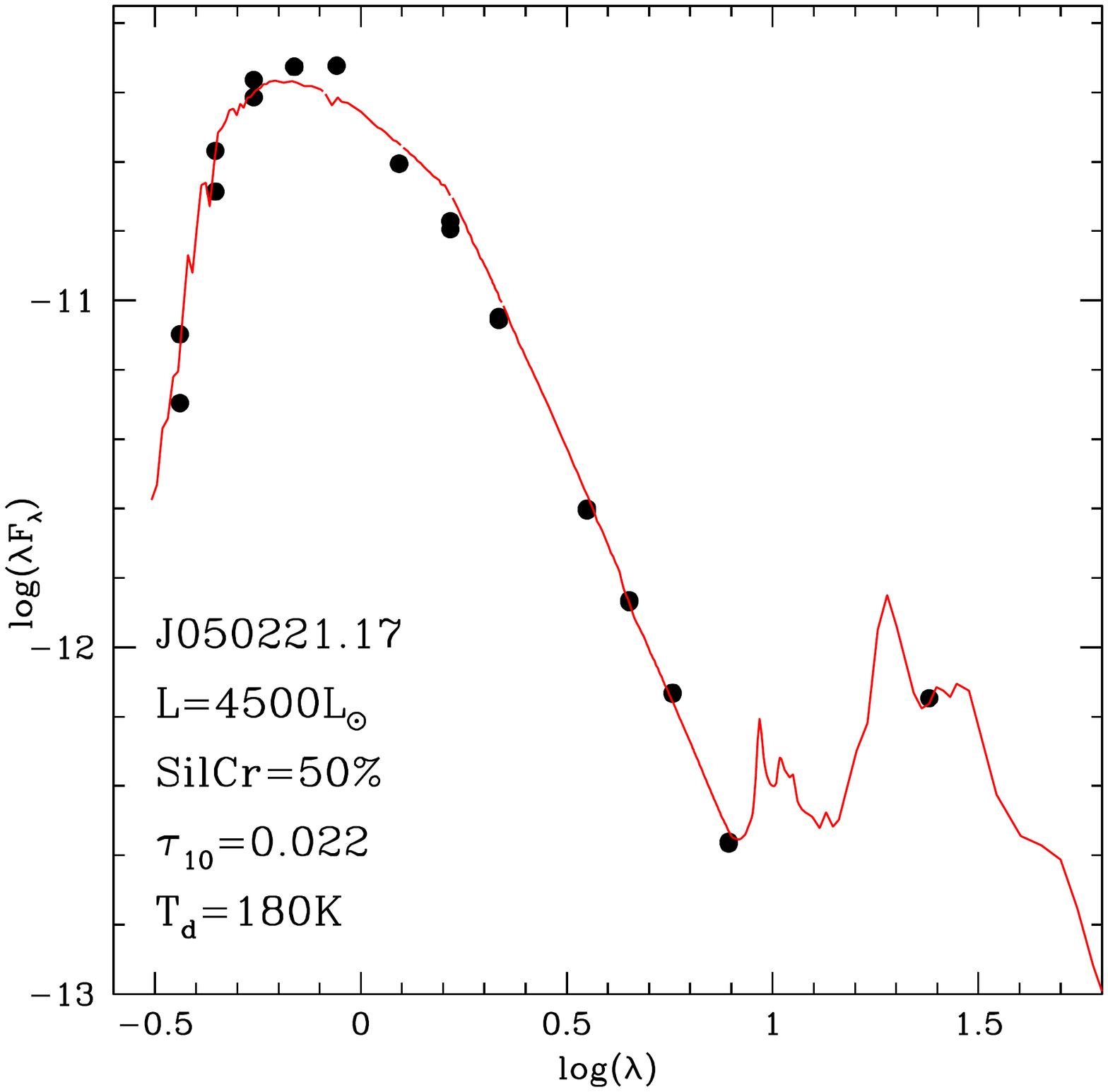}}
\end{minipage}
\begin{minipage}{0.32\textwidth}
\resizebox{1.\hsize}{!}{\includegraphics{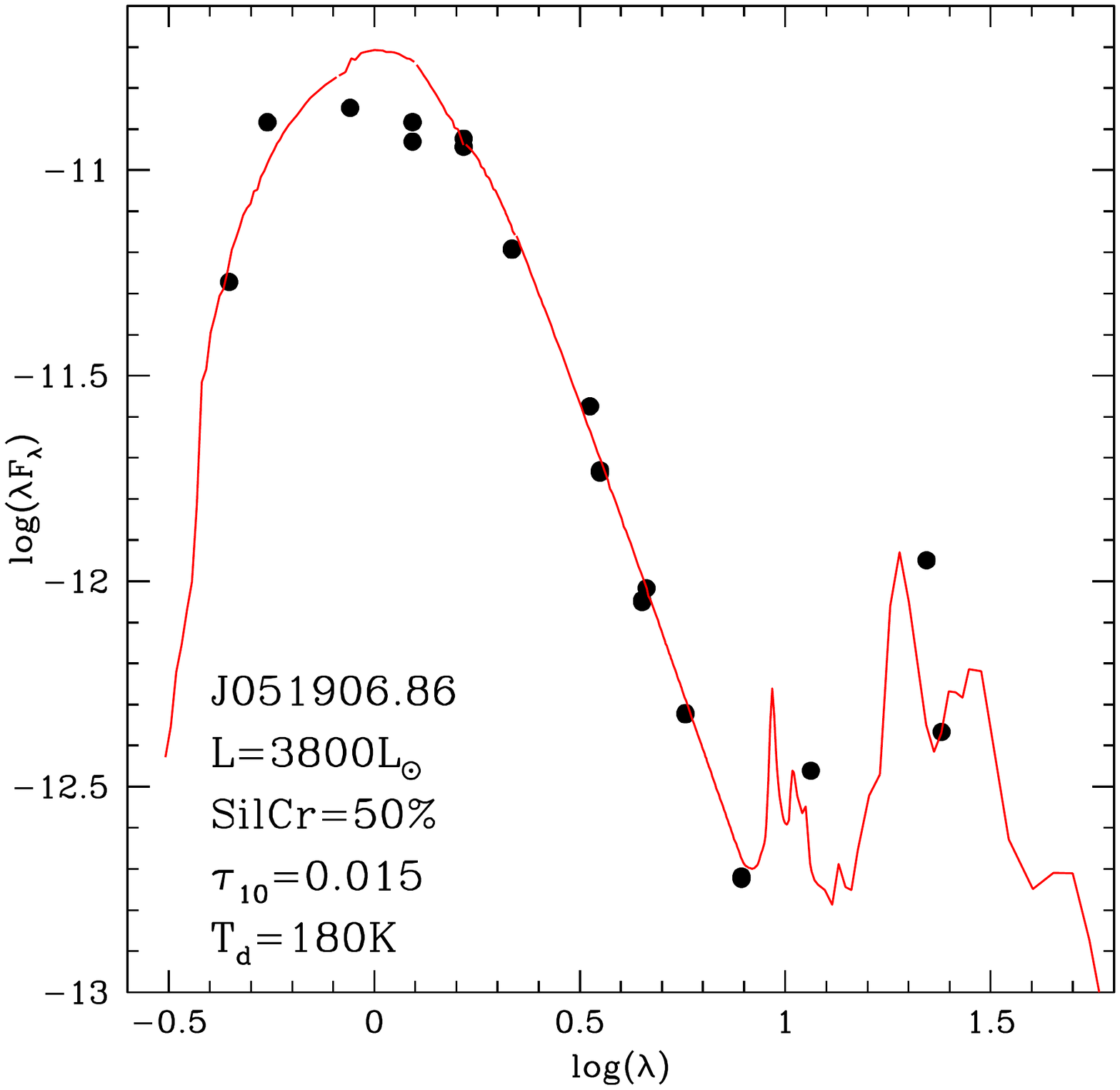}}
\end{minipage}
\vskip-40pt
\caption{Same as Fig.~\ref{fcstar}, reporting the sources that we interpret as surrounded
by silicate dust. The symbols used are the same as in Fig.~\ref{fcstar}. The fraction of silicate dust in the crystalline form is given in each panel.}
\label{fmstar}
\end{figure*}

\begin{table*}
\caption{Physical and dust properties of the LMC and SMC sources classified as shell-type
post AGB stars by K14 and K15. The quantities listed in the various columns are the following:
1 - source ID; 2, 3: metallicity and effective temperatures derived spectroscopically by 
K14 and K15; 4-7: luminosity, optical depth at $10~\mu$m, dust temperature and distance
separating the central star from the inner border of the dusty region, found via SED
fitting; 8: mass of the progenitor at the beginning of the AGB phase, deduced via
comparison of the derived luminosity with results from post-AGB evolution modelling.
}
\label{tabero}      
\centering
\addtolength{\leftskip}{-2cm}
\addtolength{\leftskip}{-2cm}
\begin{tabular}{c c c c c c c c}    
\hline      
ID & $[$Fe$/$H$]$ & ${\rm T}_{\rm eff}$[K] & ${\rm L}/{\rm L}_{\odot}$ & $\tau_{10}$ & ${\rm T}_d$[K] & ${\rm R}_{\rm in}/{\rm R}_{\odot}$ & ${\rm M}/{\rm M}_{\odot}$ \\
\hline 
J052220.98 & -0.50 & 5750 & 4500  & 0.015 & 220 & $2.63\times 10^5$ &	0.9   \\
J053250.69 & -1.10 & 6000 &	5200  & 0.006 & 250 & $2.06\times 10^5$ &	1.00  \\
J004114.10 & -1.04 & 5750 &	5800  & 0.007 & 280 & $1.65\times 10^5$ &	1.00  \\
J050632.10 & -0.40 & 7600 &	6000  & 0.003 & 240 & $2.74\times 10^5$ &	1.25  \\
J003643.94 & -0.63 & 7500 &	6500  & 0.002 & 250 & $2.56\times 10^5$ &	1.50  \\
J051848.84 & -1.00 & 6000 &	6700  & 0.008 & 240 & $2.62\times 10^5$ &	1.25  \\
J004441.03 & -1.07 & 6000 &	8500  & 0.002 & 320 & $6.64\times 10^4$ &	2.00  \\
J005803.08 & -1.03 & 6500 &	12000 & 0.025 & 280 & $1.20\times 10^5$ &	2.50  \\
\hline
J045119.94 & -0.40 & 8250 &	5000 &	0.006   & 270 & $2.97\times 10^4$ &	0.80 \\
J052241.52 & -0.50 & 8250 &	5000 &	0.017	& 400 &	$2.26\times 10^4$ &	0.80 \\ 
J052740.75 & -0.50 & 8250 &	5200 &	0.024	& 470 &	$1.22\times 10^4$ &	0.90 \\
J050221.17 & -0.60 & 5250 &	4500 &	0.022	& 180 &	$5.45\times 10^4$ &	0.80 \\ 
J051906.86 & -1.30 & 5500 &	3800 &	0.015	& 180 &	$5.13\times 10^4$ &	0.65 \\
\hline
\label{tabpost}
\end{tabular}
\end{table*}

Fig.~\ref{fcstar} and \ref{fmstar} show the analysis done for each of the sources
considered, divided according to whether they are interpreted as surrounded by carbon dust 
and silicates, respectively. We note that the error bars of the photometric data of the post-AGBs 
in SMC/LMC (from K14 and K15) are small and hence we do not display them in our SED fits.
In the cases where multi-epoch observations for the same band are available, we show all
the results in a vertical sequence, in conjunction with the central wavelength of the
corresponding filter.

Tab.~\ref{tabero} reports the metallicities, the effective 
temperatures, and the quantities derived from SED fitting, namely the luminosity and
the dust properties, i.e. the mineralogy, the optical depth, the dust temperature and
the distance from the centre of the star of the inner border of the dust layer. The
last column gives the estimated mass of the progenitors at the beginning 
of the AGB phase (see next section).

 We find that the dust in the surroundings of the stars that we classify as oxygen-rich
sources is composed by a mixture of amorphous and crystalline silicates, with the
latter component accounting for $\sim 50\%$ of the total dust. The presence of
a significant fraction of dust in the crystalline form is required 
to reproduce the strongly negative slope of the SED in
the $\lambda > 20~\mu$m spectral region, as determined by the combination of the
WISE4 and MIPS data. Use of a pure amorphous dust in the SED fitting procedure does not
allow to reproduce the shape of the SED, independently of the optical constants adopted.

We note that this conclusion relies on the assumptions upon which the DUSTY code
is based, namely that the dusty zone is spherically distributed and is characterized by a single 
dust component and temperature. Taking into account all these factors would not
change the general conclusions drawn regarding the progenitors of the individual
sources and the amount of dust in their surroundings, but would likely reflect into
the relative fractions of amorphous and crystalline silicates in the case of M-type 
sources.

The presence of crystalline silicate dust has been so far detected in AGB sources with
optically thick dust envelopes in the Galaxy \citep{sylvester99}. Furthermore, 
emission bands associated to crystalline silicates in evolved stars in the LMC were 
studied by \citet{jones14}, down to (gas) mass-loss rates of $\sim 10^{-6}~{\rm M}_{\odot}/$yr,
fairly consistent with the mass-loss rates that low-luminosity, oxygen-rich stars attain
at the tip of the AGB (see section \ref{lowm} below). The explanation proposed here holds as 
far as no change in the structure of silicate dust occurs as the dust moves away while the star 
contract to the post-AGB phase.

\section{The progenitors of the post-AGB stars sample}
\label{disc}

We characterize the sources discussed in the previous section to infer the mass and
formation epoch of the progenitors. To this aim we consider the position of the stars
on the HR diagram, based on the values of effective temperature and luminosity listed in
Tab.~\ref{tabero}, and the evolutionary tracks of model stars of different mass, with 
the metallicity similar to those reported in the col.~2 of Tab.~\ref{tabero}. This
approach can be safely used in the case of post-AGBs, given the peculiar morphology
of the tracks, which run practically horizontal on the HR plane. 

\begin{figure*}
\begin{minipage}{0.48\textwidth}
\resizebox{1.\hsize}{!}{\includegraphics{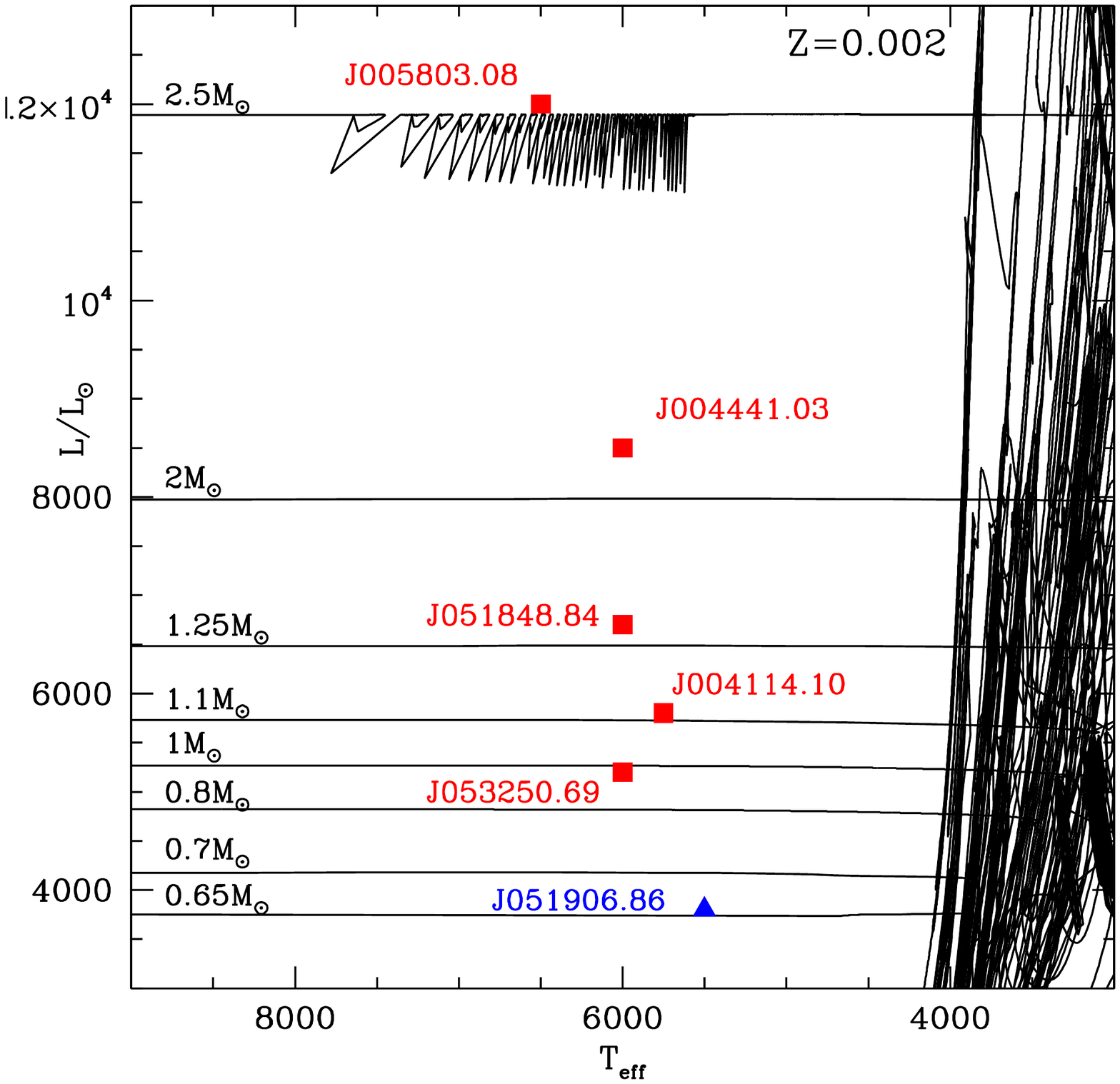}}
\end{minipage}
\begin{minipage}{0.48\textwidth}
\resizebox{1.\hsize}{!}{\includegraphics{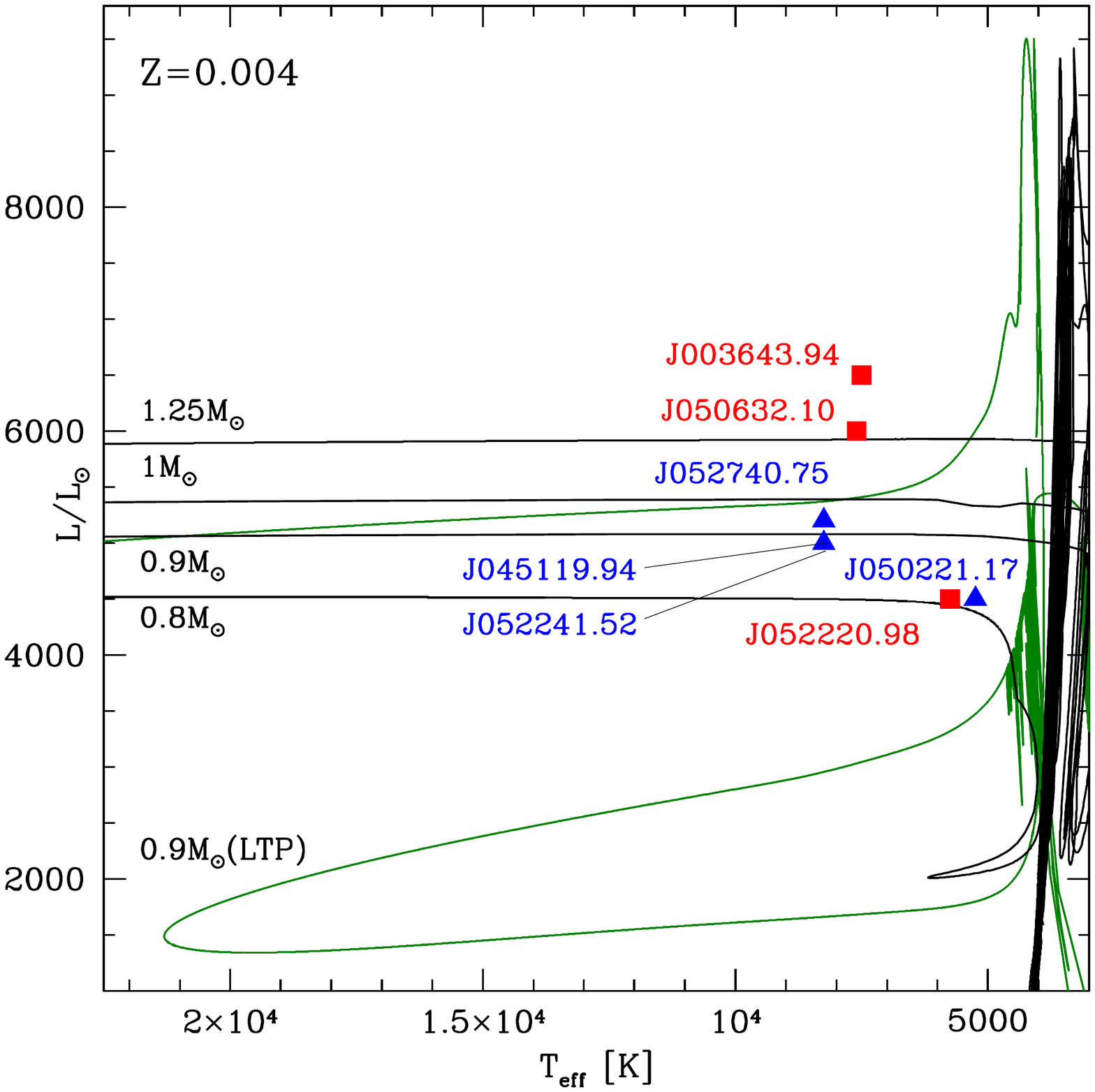}}
\end{minipage}
\vskip-60pt
\caption{Post-AGB evolutionary tracks of model stars of metallicity $Z=2\times 10^{-3}$ (left panel)
and $Z=4\times 10^{-3}$ (right). The green track in the right panel refers to a $0.9~{\rm M}_{\odot}$
model star, which experienced a late TP, shortly after the beginning of the contraction towards
the post-AGB. Coloured points indicate the observationally derived position of the sources 
considered in the present investigation, according to the effective temperatures given 
in K14 and K15 and the
luminosities derived in section \ref{sedfit}. Red squares and blue triangles refer to the sources 
surrounded by carbonaceous dust and silicates, respectively.}
\label{ftracks}
\end{figure*}

We show in Fig.~\ref{ftracks} the position of the different sources on the HR diagram, 
overimposed to the evolutionary tracks of $0.65-2.5~{\rm M}_{\odot}$ stars\footnote{The 
masses considered here refer to the beginning of the AGB phase, thus they
do not take into account possible mass loss during the RGB ascending}. We separate metal-poor 
sources, with $[$Fe$/$H$]\sim -1$, from the higher metallicity counterparts. The former, 
reported in the left panel of Fig.~\ref{ftracks}, are compared with tracks of metallicity
$Z=2\times 10^{-3}$, whereas to interpret the latter group we use evolutionary sequences 
with metallicity $Z=4\times 10^{-3}$. 

In the following we present the interpretation of the stars, discussing separately the carbon stars
and the oxygen-rich objects, grouped according to the estimated age and metallicity. When available,
results from high-resolution spectroscopy, which allowed the determination of the surface chemical 
composition, will be compared with the expectations from stellar evolution modelling, thus
providing a further test of the interpretation proposed here.

\subsection{Carbon stars}
\subsubsection{J053250.69 and J004114.10: low-mass, metal-poor carbon stars}
\label{lowmc}

\begin{figure*}
\begin{minipage}{0.48\textwidth}
\resizebox{1.\hsize}{!}{\includegraphics{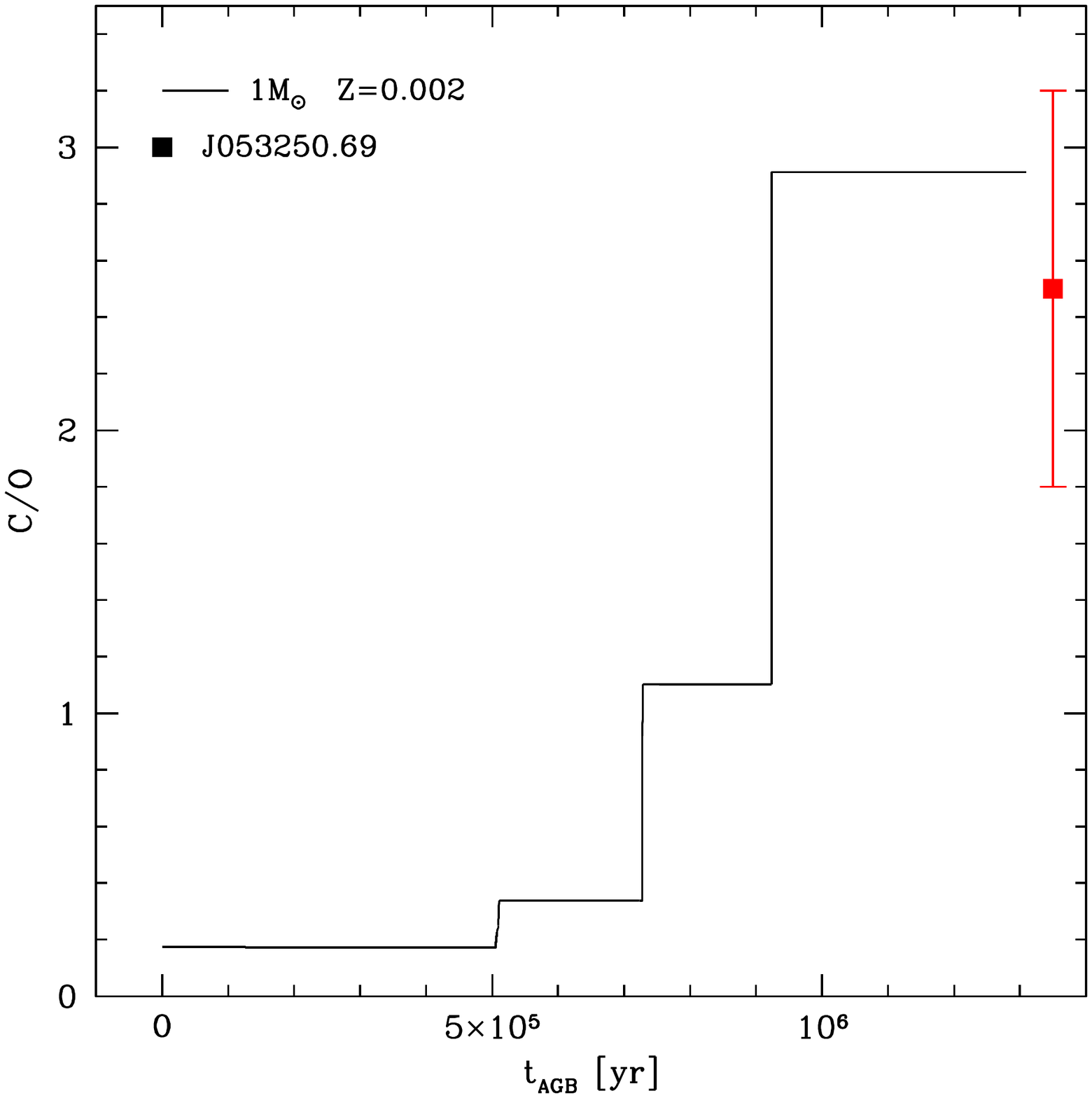}}
\end{minipage}
\begin{minipage}{0.48\textwidth}
\resizebox{1.\hsize}{!}{\includegraphics{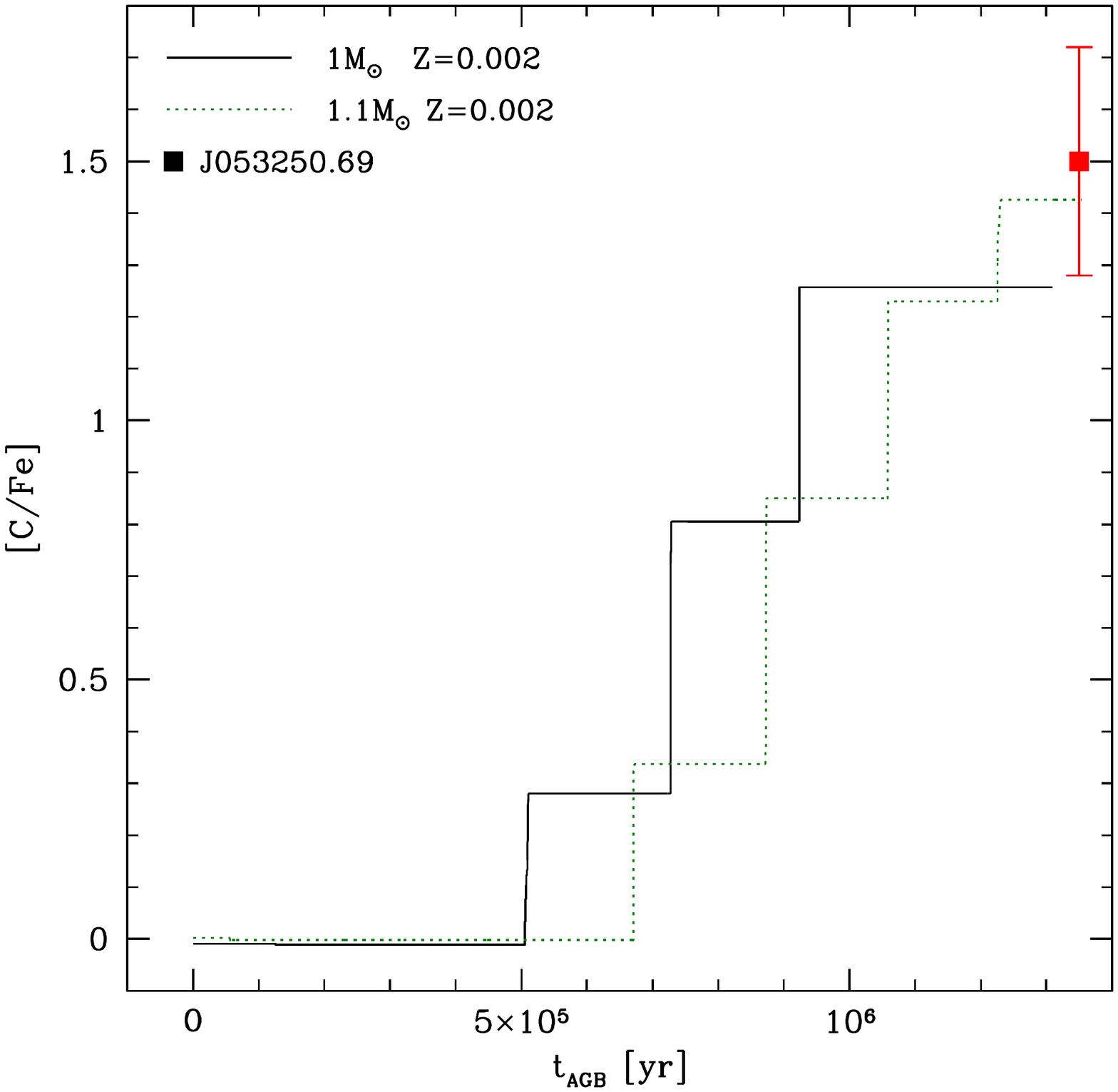}}
\end{minipage}
\vskip-60pt
\caption{{\bf Left}: Time variation of the surface C$/$O ratio of a $1~{\rm M}_{\odot}$, metal-poor model star
during the AGB phase, until the post-AGB evolution. Red square and line indicate the measured value
and the error bar from \citet{vanaarle13}. {\bf Right}: The time evolution of the surface carbon of two model
stars of mass $1~{\rm M}_{\odot}$ (solid, black line) and $1.1~{\rm M}_{\odot}$ (green, dotted line),
compared with results from \citet{vanaarle13}.}
\label{fg1}
\end{figure*}

The lowest luminosity stars among the metal-poor objects surrounded by carbon dust are J053250.69 and 
J004114.10. From their position on the HR diagram, shown in the left panel of Fig.~\ref{ftracks}, we deduce that
they descend from stars with mass equal to or slightly greater than $1~{\rm M}_{\odot}$. As far as J053250.69 is 
concerned, the presence of carbon dust is consistent with the results
from \citet{vanaarle13}, that indicate that the surface chemistry is enriched in carbon. The surface
C$/$O and $[$C$/$Fe$]$ are in agreement with results from stellar evolution modelling: this is shown
in Fig.~\ref{fg1}, where the values given in \citet{vanaarle13} are compared with the AGB time variation of
the C$/$O ratio and of the surface carbon content of a $1~{\rm M}_{\odot}$ model star. The expected increase 
in the surface $[$O$/$Fe$]$ (not shown in Fig.~\ref{fg1}), of the order of $0.1$ dex, is $0.1-0.2$ dex below 
the lower limit given in \citet{vanaarle13}.

Regarding J004114.10, there are no results from spectroscopy. According to the estimated luminosity we expect that
the progenitor is $\sim 0.1~{\rm M}_{\odot}$ more massive than for J053250.69. In this case we expect a higher
carbon enrichment, with a final $[$C$/$Fe$] \sim 1.5$ (see right panel of Fig.~\ref{fg1}).

Based on these results, and considering a $0.1-0.2~{\rm M}_{\odot}$ mass loss during the red-giant branch (RGB) ascending,
we expect that these two sources descend from metal-poor progenitors of mass in the $1.2-1.3~{\rm M}_{\odot}$
range, born $\sim 3$ Gyr ago.

\begin{figure}
\resizebox{1.\hsize}{!}{\includegraphics{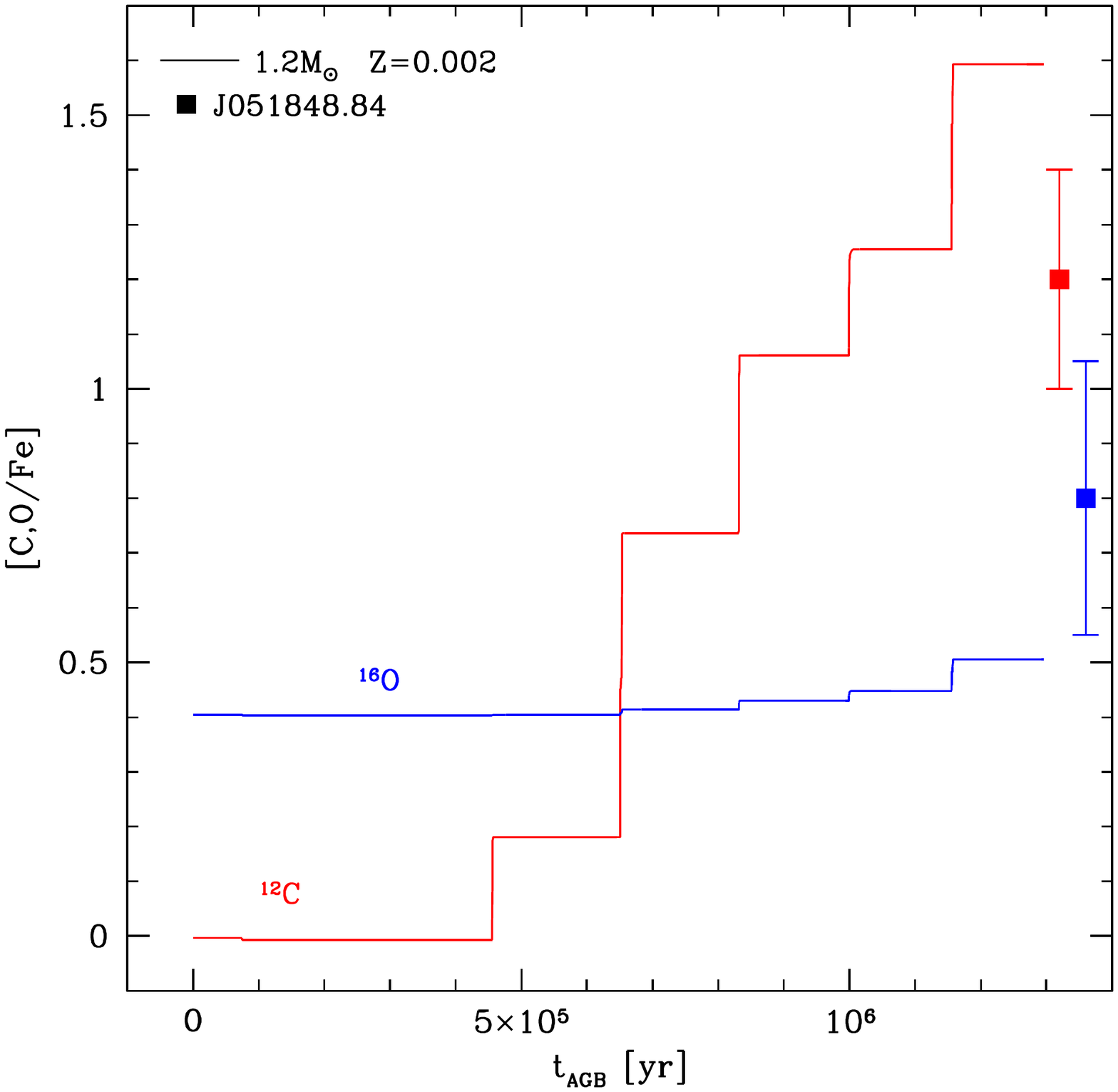}}
\vskip-60pt
\caption{Time variation of the surface carbon (red line) and oxygen (blue) content
during the AGB evolution of a $1.25~{\rm M}_{\odot}$ model star of metallicity
$Z=0.002$. The red and blue points on the right side of the plane (with the
corresponding error bars) indicate results for J051848.84 from \citet{desmedt15}.}
\label{fg2}
\end{figure}

\subsubsection{J051848.84}
For this source we derive a luminosity ${\rm L} \sim 6700~{\rm L}_{\odot}$, consistent with the
conclusions from \citet{desmedt15}. This luminosity, as shown in the left panel of Fig.~\ref{ftracks},
corresponds to a $\sim 1.25~{\rm M}_{\odot}$ star.
The conclusion drawn in section \ref{sedfit}, that the star is surrounded by
carbon dust, is consistent with the discussion in \citet{desmedt15}, that this
source is enriched in carbon, with $[$C$/$Fe$]=1.2$. The s-process enrichment claimed by 
\citet{desmedt15} is a further hint of the action of repeated TDU events. 
The observed surface $^{12}$C and $^{16}$O are in agreement with results
from stellar evolution modelling, as shown in Fig.~\ref{fg2}, where the results from 
\citet{desmedt15} are compared with the time evolution of the surface carbon and oxygen of
a $1.25~{\rm M}_{\odot}$ model star. Assuming a $0.1~{\rm M}_{\odot}$ mass loss during the
RGB, we deduce that the progenitor of J051848.84 was a $1.3-1.4{\rm M}_{\odot}$ star,
formed $\sim 2.5$ Gyr ago.

\begin{figure}
\resizebox{1.\hsize}{!}{\includegraphics{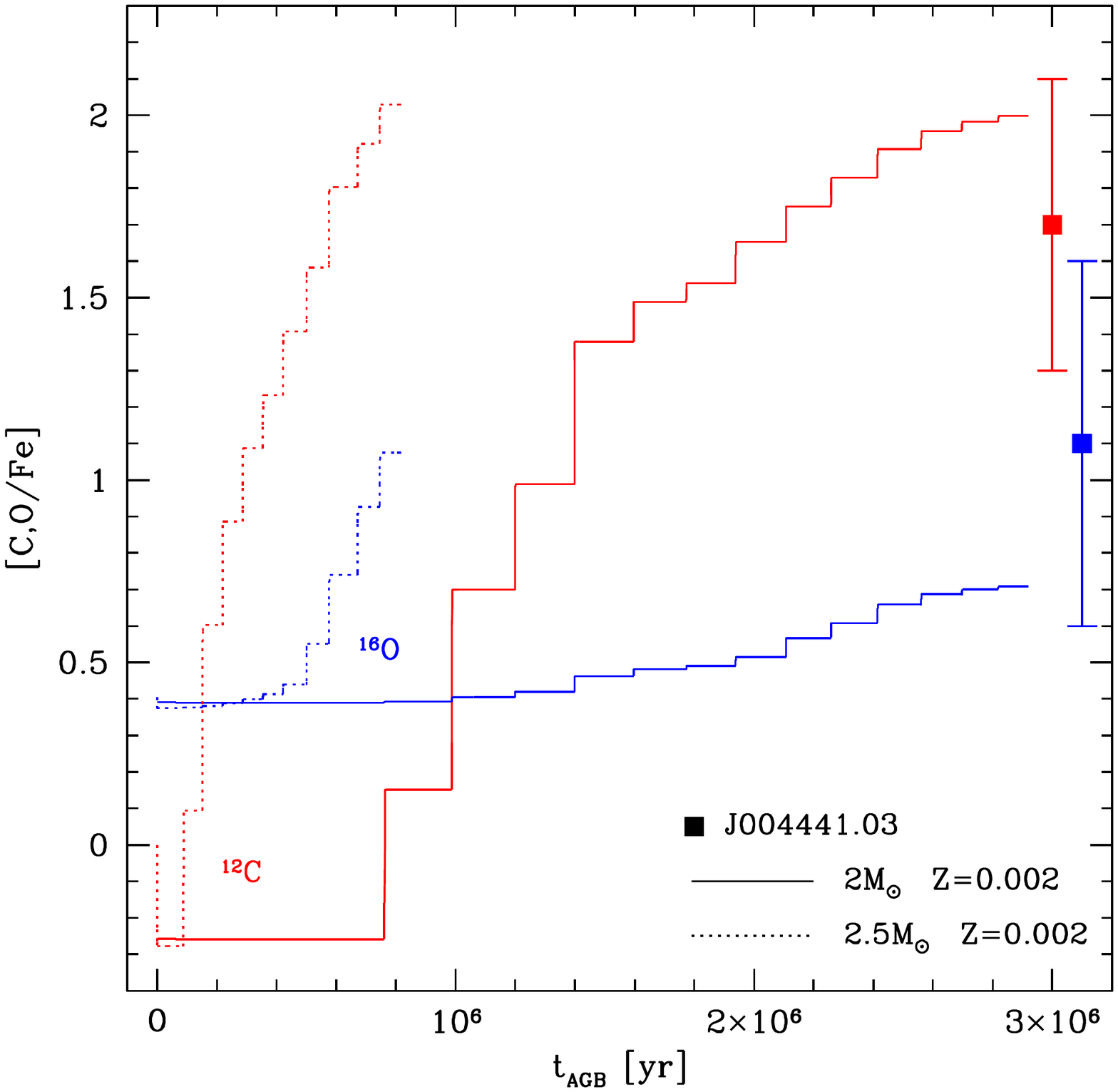}}
\vskip-60pt
\caption{Time variation of the surface carbon (red lines) and oxygen (blue) content
during the AGB evolution of two $Z=0.002$ model stars of mass $2~{\rm M}_{\odot}$ 
(solid lines) and $2.5~{\rm M}_{\odot}$  (dotted). The red and blue points on the 
right side of the plane (with the corresponding error bars) indicate results for J004441.03 
from \citet{desmedt15}}
\label{fg3}
\end{figure}

\subsubsection{J004441.03}
This source, with an estimated luminosity of $8500~{\rm L}_{\odot}$, is among the
brightest objects in the sample considered in the present investigation. We concluded
in the previous section that it is surrounded by carbon dust: this is consistent both 
with the results from \citet{desmedt15}, which indicate significant carbon and s-process enrichment, 
and with the interpretation based on the position on the HR diagram (see Fig.~\ref{ftracks}),
which suggests a $\sim 2~{\rm M}_{\odot}$ progenitor. According to this understanding 
this is the first source, among those discussed so far, which did not experience the helium 
flash, rather core helium burning occurred in quiescent conditions of thermal stability.
Stars of mass in the $2-3~{\rm M}_{\odot}$ range are expected to experience a 
series of TDU events that trigger a significant surface carbon enrichment: 
\citet{flavia15a, flavia15b} concluded that the sources in the LMC exhibiting
the largest infrared excesses are the progeny of $2-3~{\rm M}_{\odot}$ stars.

This interpretation finds a confirmation in the results given in \citet{desmedt15}, who 
find an extremely large surface carbon mass fraction $[$C$/$Fe$] \sim 1.7$. 
The results shown in Fig.\ref{fg3} confirm the robustness of the present analysis, as the
final surface content of both carbon and oxygen are fully consistent with the values given 
by \citet{desmedt15}. The estimated age of J004441.03 is $\sim 1$ Gyr.

\subsubsection{J005803.08}
\label{bright}
This is the brightest star in the sample, with an estimated luminosity 
$\sim 12000~{\rm L}_{\odot}$. The comparison with results from post-AGB
evolution modelling, shown in the left panel of Fig.~\ref{ftracks}, suggests that
the progenitor of this source is a $2.5~{\rm M}_{\odot}$ star. The expected
variation of the surface chemistry of such a star, in terms of the surface
$^{12}$C and $^{16}$O mass fractions, is shown in Fig.~\ref{fg3}, with
dotted lines. 

Similarly to the star examined previously, we expect that this source experienced
several TDU events, that caused a significant increase in the surface carbon,
with a final $[$C$/$Fe$]>2$. Some oxygen enrichment is also expected
(see Fig.~\ref{fg3}), as also the increase in the s-process abundance.


J005803.08 is highest mass among the sources analysed here, which formed
most recently, with an estimated age slightly older than half Gyr.

\begin{figure}
\resizebox{1.\hsize}{!}{\includegraphics{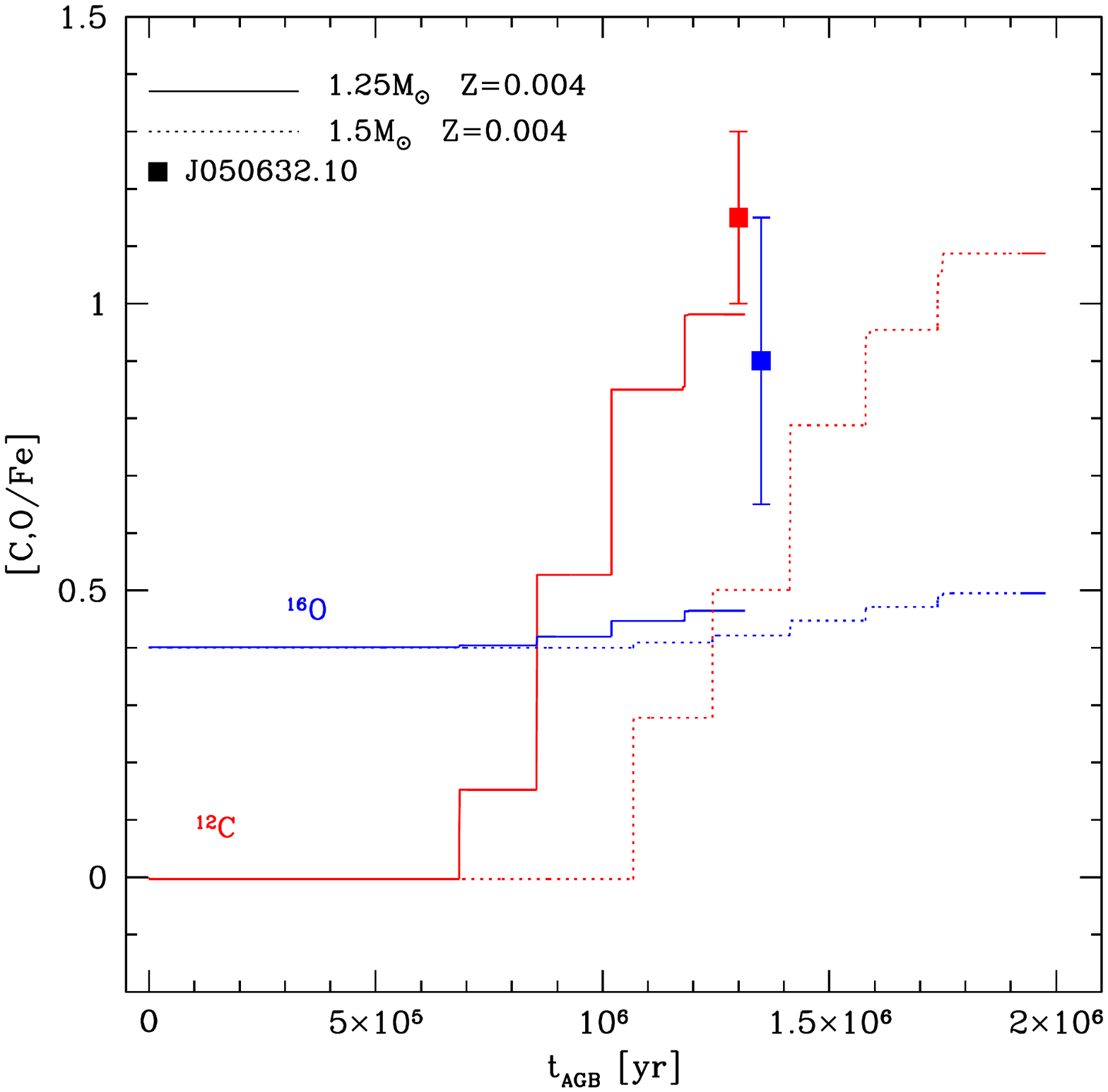}}
\vskip-60pt
\caption{Time variation of the surface carbon (red lines) and oxygen (blue) content
during the AGB evolution of two $Z=0.004$ model stars of mass $1.25~{\rm M}_{\odot}$ 
(solid lines) and $1.5~{\rm M}_{\odot}$ (dotted). The red and blue points in the middle
of the plane (with the corresponding error bars) indicate results for J050632.10 
from \citet{vanaarle13}}
\label{fg4}
\end{figure}

\subsubsection{J050632.10 and J003643.94}
\label{z4m3}
From SED fitting (see Fig.~\ref{fcstar}) we find that both these sources with sub-solar 
chemical composition are surrounded by carbon dust. This is consistent with the estimated 
luminosities, in the $6000-6500~{\rm L}_{\odot}$ range, which based on post-AGB modelling 
correspond to progenitors of mass slightly above the solar mass (see right panel of
Fig.~\ref{ftracks}), which are expected to experience a  sufficiently large number of TDU 
episodes to reach the carbon star stage.

\citet{vanaarle13} presented results from spectroscopy for J050632.10, which confirmed 
carbon enrichment, with $[$C$/$Fe$]$ slightly above unity. As shown in Fig.~\ref{fg4},
this is consistent with the expected carbon enrichment of $1.25~{\rm M}_{\odot}$ stars of 
metallicity $Z=0.004$, which is the derived mass of the progenitor of this source, based on the
luminosity. On the other hand, the expected oxygen enrichment is smaller
than indicated by \citet{vanaarle13}, even in the hypothesis that the gas from
which the star formed was significantly $\alpha-$enhanced, with $[\alpha/{\rm Fe}]=+0.4$.

J003643.94 is slightly brighter than J050632.10 (see right panel of Fig.~\ref{ftracks}),
thus we deduce a higher mass progenitor, with ${\rm M} \sim 1.5~{\rm M}_{\odot}$. The
results shown in Fig.~\ref{fg4}, where the variation of the surface $^{12}$C and $^{16}$O
for the $1.25~{\rm M}_{\odot}$ and $1.5~{\rm M}_{\odot}$ cases are reported, indicate that 
the final surface abundances of carbon and oxygen
should be pretty similar to those found for J050632.10

According to the present interpretation the formation of J003643.94 occurred
around 2 Gyr ago, whereas J050632.10 is older, with an estimated age slightly
above 3 Gyr.

\subsubsection{J052220.98: a (too) faint carbon star}
\label{faintc}
The position of J052220.98 on the HR diagram reported in the right panel of Fig.~\ref{ftracks} 
suggests a $\sim 0.8~{\rm M}_{\odot}$ progenitor, an interpretation that 
however is at odds with the analysis done in section \ref{sedfit}, where
we found that this source is surrounded by carbon dust. Indeed, as discussed earlier in
this section, we expect that carbon stars of metallicity similar to J052220.98 evolve 
to the post-AGB phase with luminosities above $\sim 5000~{\rm L}_{\odot}$, which translates 
into ${\rm M} > 0.9~{\rm M}_{\odot}$ progenitors. 

A possible explanation for this apparently anomalous behaviour is that J052220.98
descends from a low mass progenitor, which experienced an extraordinary deep TDU event 
during the late evolutionary phases: the consequent large carbon enrichment would
have turned the star into a carbon star. While this possibility cannot be ruled out,
we are more predisposed to consider the hypothesis that this source experienced
a late TP shortly after the beginning of the contraction towards the post-AGB.
The idea that helium ignition might occur after the stars leave the AGB was
proposed by \citet{iben83b}, then further investigated by \citet{blocker01},
who studied different cases, distinguished by the time after the beginning
of the overall contraction when the late TP is ignited. In particular, 
in the LTP case discussed by \citet{blocker01}, the evolutionary tracks 
are characterized by a first excursion to the blue, followed by the expansion to
the red, before the standard post-AGB contraction begins. An example of such a behaviour 
is reported in the right panel of Fig.~\ref{ftracks}, in which the green line refers to 
the evolution of a $0.9~{\rm M}_{\odot}$ model star, that underwent helium ignition 
immediately after the start of the overall contraction.

During the occurrence of the loop the luminosity of the star is fainter than expected 
on the basis of the classic core mass - luminosity relationship. We therefore
suggest that J052220.98 descends from a $\sim 0.9-1~{\rm M}_{\odot}$ progenitor,
which reached the C-star stage during the AGB, and is now evolving through a loop
similar to the one shown in the figure, after having experienced a late TP.

\subsection{Oxygen-rich stars}
\label{orich}
Five out of the sources investigated here show evidence of silicate dust in their surroundings,
based on the analysis presented in section \ref{sedfit}. These stars are indicated with
blue triangles in Fig.~\ref{ftracks}.

A first possibility to explain these objects is that they descend from ${\rm M} > 3~{\rm M}_{\odot}$
stars, which experienced HBB during the AGB phase; the latter mechanism prevents the surface carbon 
enrichment, owing to the proton-capture activity experienced at the base of the convective envelope, 
that depletes the surface $^{12}$C. This option can be safely ruled out in the present case, as
such massive AGBs are expected to enter the post-AGB phase with luminosities not below
$20000~{\rm L}_{\odot}$, thus significantly higher than those of the stars considered here,
as clear in Fig.\ref{ftracks}.

An alternative explanation is that these sources are the progeny of low-mass 
stars, that failed to reach the C-star stage. For a given metallicity a
luminosity threshold exists, separating carbon stars from their lower mass
(and fainter) counterparts, that evolve as oxygen-rich objects: indeed the 
latter stars at the beginning of
the AGB phase are composed by a $\sim 0.5~{\rm M}_{\odot}$ degenerate core 
and by a convective envelope of a few tenths of solar masses, that is lost
via stellar winds before the number of TDUs required to reach the
carbon star stage is experienced. The luminosity threshold
is sensitive to the metallicity, as the achievement of the C-star stage
is easier the lower the metallicity, owing to the smaller amount of
oxygen in the envelope of the star. In the $Z=0.002$ case we find that
carbon stars should have ${\rm L}>4700~{\rm L}_{\odot}$, whereas for
$Z=0.004$ the threshold luminosity is $\sim 5200~{\rm L}_{\odot}$.

The luminosities of the five sources characterized by the presence of silicate
dust in their surroundings are within the limits given above, as shown in
Fig.~\ref{ftracks}. In particular J051906.86 and J050221.17 are significantly
fainter than the threshold luminosities discussed above, thus they are
interpreted as the progeny of low-mass stars, with mass at the beginning
of the AGB phase of $\sim 0.7~{\rm M}_{\odot}$ (J051906.86) and
$\sim 0.8~{\rm M}_{\odot}$ (J050221.17), which, by assuming typical RGB mass
loss $0.1-0.2~{\rm M}_{\odot}$, correspond to ages in the $8-10$ Gyr range.
These are the oldest objects in the sample investigated here.

On the other hand J052740.75, J045119.94 and J052241.52 are slightly
brighter, with luminosities close to the afore discussed threshold, holding
for the $Z=0.004$ population. These luminosities correspond to
$\sim 0.85~{\rm M}_{\odot}$ progenitors, that formed between 6 and
8 Gyr ago. We might expect to observe some carbon and s-process enrichment
at their surface, as it is likely that they experienced some TDU events before
the entire envelope was lost.

\begin{figure}
\resizebox{1.\hsize}{!}{\includegraphics{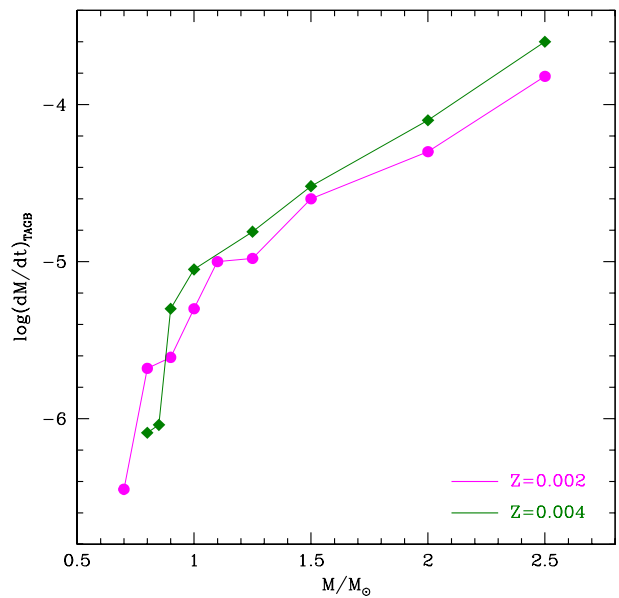}}
\vskip-60pt
\caption{Mass-loss rate at the tip of the AGB for stars of different
mass, reported on the abscissa, and metallicity $Z=0.002$ (magenta points) and
$Z=0.004$ (green diamonds). The values reported in the plane were calculated
by stellar evolution modelling.}
\label{fmlo}
\end{figure}

\begin{figure*}
\begin{minipage}{0.48\textwidth}
\resizebox{1.\hsize}{!}{\includegraphics{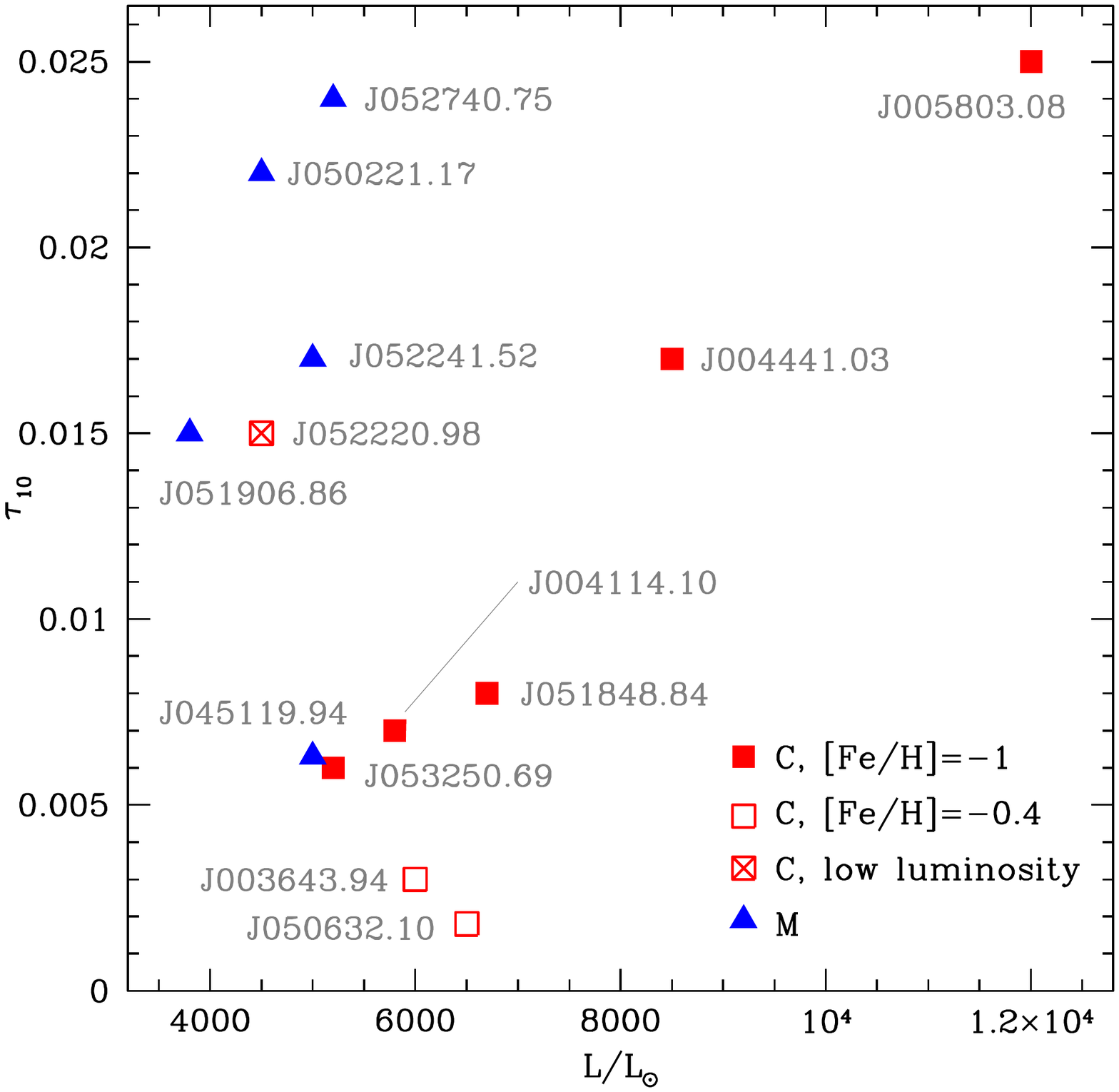}}
\end{minipage}
\begin{minipage}{0.48\textwidth}
\resizebox{1.\hsize}{!}{\includegraphics{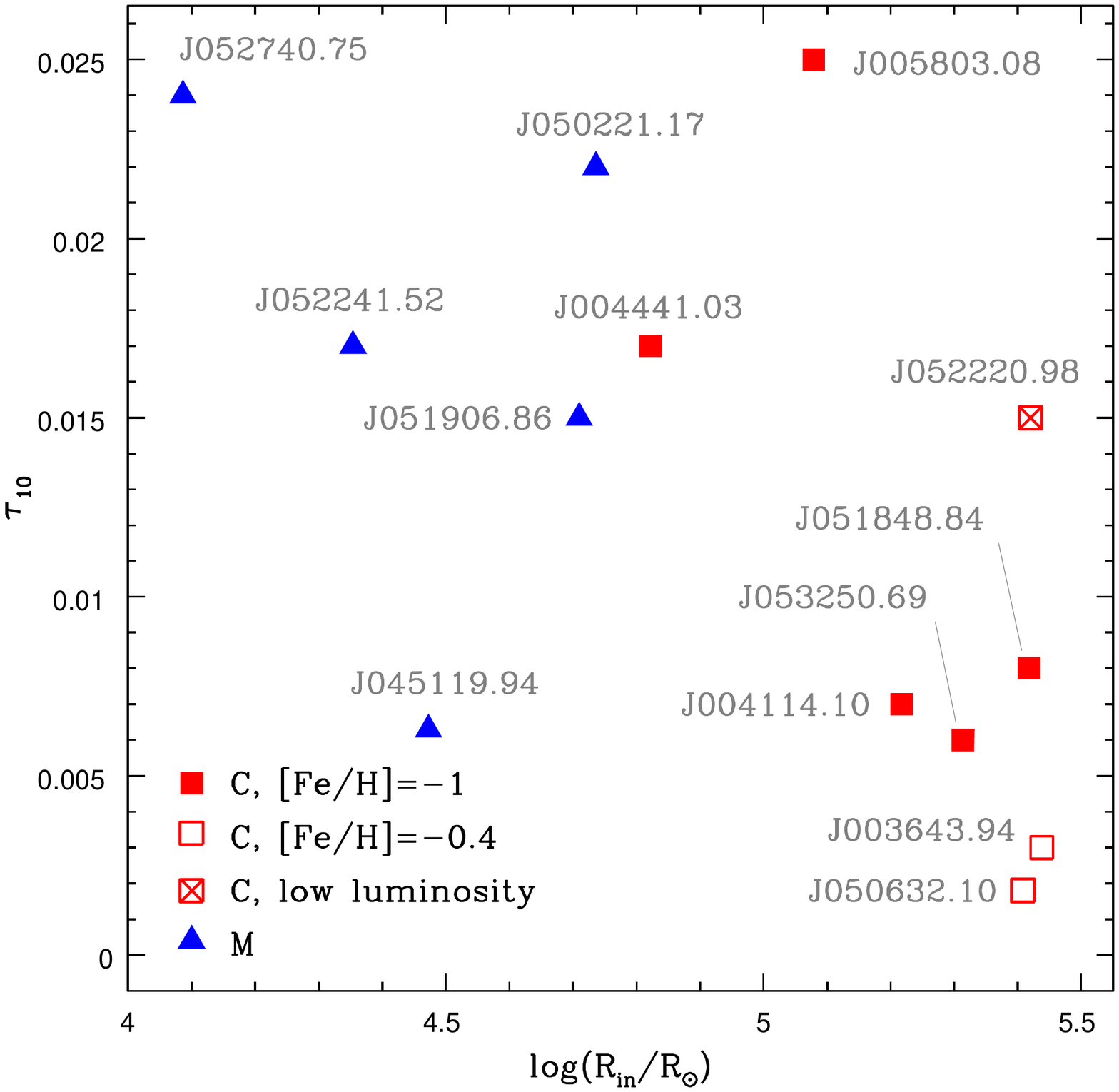}}
\end{minipage}
\vskip-60pt
\caption{The optical depth of the different sources obtained from SED fitting in section 
\ref{sedfit} as a function of the luminosity of the stars (left panel) and of the distance
of the internal border of the dusty region from the center of the star (right). 
Red squares and blue triangles indicates stars surrounded by carbonaceous dust
and silicates, respectively. Crossed red squares refer to stars exhibiting evidence
of carbon dust, despite their luminosities are consistent with an M-type nature.
}
\label{fall}
\end{figure*}

\section{Dust production during the late AGB phases}
\label{lateagb}
During the AGB evolution the stars gradually lose the external convective mantle.
While the central regions contract, the external layers become
less and less bound on the gravitational side to the compact core and expand, and
the mass loss rate is consequently enhanced. This general rule holds for low-mass objects
that evolve as M-type stars, such as those discussed in section \ref{orich}, 
and even more for carbon stars: indeed in the latter
case the expansion of the structure is further favoured by the rise in the
surface carbon, which is accompanied by a significant increase in the 
molecular opacities \citep{marigo02}. Massive AGB stars that experience HBB follow a different
behaviour, as in this case the largest mass loss rates are experienced during the
phases when HBB is strongest \citep{ventura15}, which happens before a significant 
fraction of the
envelope is lost. However this is not relevant in the present context, as we showed in
section \ref{disc} that the post-AGB stars discussed here descend 
from ${\rm M} < 3~{\rm M}_{\odot}$ progenitors.

The above arguments indicate that the final AGB phases are the most relevant
to understand dust production from stars. This is because most of the mass is lost 
during the last interpulse phases and the dust production rate, tightly correlated with
$\dot{\rm M}$, increases rapidly towards the end of the AGB evolution.
Understanding dust production during the tip of the AGB (TAGB), i.e. the phase preceding the beginning
of the contraction of the star towards the  post-AGB,
is therefore crucial to evaluate the dust yields from single stars and
the overall contribution from these objects to dust production in the host system. 
Indeed the total dust production 
rate of some galaxies is mostly determined by a low fraction of AGB stars, which are
currently evolving through the very final AGB phases \citep{flavia16, flavia18, flavia19}. 
In addition, the comprehension of the TAGB turns crucial to interpret the 
distribution of stars of galaxies in the observational colour-colour and colour-magnitude 
planes, because it is while evolving thorugh the TAGB that the stars are expected
to exhibit the largest infrared excesses, thus the reddest IR
colours \citep{ester21, flavia21}.

Fig.~\ref{fmlo} shows the mass-loss rate at the TAGB, $\dot{\rm M}^{\rm TAGB}$, calculated 
via stellar evolution modelling, as a function of the mass of the star at the 
beginning of the AGB phase. $\dot{\rm M}^{\rm TAGB}$ increases with stellar mass, as
higher mass stars reach higher luminosities and radii. It is clear in the figure
the steep rise in $\dot{\rm M}^{\rm TAGB}$ in correspondence of the threshold mass at 
which transition from M-type to C-stars takes place, which is connected to the 
afore discussed increase in the surface molecular opacities, taking place when C-rich 
molecules form.

In the following section we will critically evaluate the mass loss rates
reported in Fig.~\ref{fmlo}, by comparing the theoretical expectations regarding the
dust in the surroundings of post-AGB stars with the results discussed in section 
\ref{sedfit} and \ref{disc}. 

On this regard we will consider the characterization of the individual sources presented in
section \ref{disc}, which allows the derivation of the mass of the progenitors and,
based on the results shown in Fig.~\ref{fmlo}, of $\dot{\rm M}^{\rm TAGB}$. These rates, 
properly scaled with the effective temperature, lead to the determination of the variation of 
the mass loss rates experienced by the stars during the contraction to the post-AGB.

The comparison with the results deduced in section \ref{sedfit}, particularly of the derived
values of $\tau_{10}$, allows to test the $\dot{\rm M}^{\rm TAGB}$ values reported in Fig.~\ref{fmlo} and, 
at the same time, to identify the evolutionary phase when the dust responsible for the currently 
observed IR excess was released. To this scope we will concentrate on some specific points
distributed along the post-AGB evolutionary tracks, each identified by the value of the effective 
temperature. For each of these points we will derive $\tau_{10}^{\rm onset}$, the optical depth
that characterizes the SED of the star if it was observed during the evolutionary stage
considered. The values of $\tau_{10}^{\rm onset}$ are calculated on the basis of stellar evolution + dust 
formation modelling, and the application of Eq.~3.

We note that the possibility that dust production continues after the beginning of the
contraction to the post-AGB was already considered, e.g. by \citet{vanhoof97}, who explored
both the possibilities that dust formation ceased after the AGB, or that it carries on during
the post-AGB phase. In the following we will show that for all the sources considered the observations 
are non consistent with the possibility that the dust was released after the effective temperatures
raises above 4000 K. Therefore, here we do not explore dust formation during the post-AGB phase
through which the sources are evolving nowadays, rather we consider the possibilities that 
formation of dust was halted either at the tip of the AGB or during the early phases after
the beginning of the contraction to the post-AGB.

The second step will consist in the determination of $\tau_{10}^{\rm now}$, the optical depth that 
characterizes the current SED of the star, if the dust was released during the stages for which 
$\tau_{10}^{\rm onset}$ was calculated. To this aim we 
consider the general expression given by Eq.~3. In the computation of the integral 
on the right-hand side most of the contribution comes from the regions close to the 
inner border of the dusty layer, 
as $n_d$ decreases with distance: therefore, $\tau_{10}$ is mainly determined by the
thermodynamic conditions, particularly the density, at the distance ${\rm R}_{in}$ from the
centre of the star. Assuming that $n_d$ scales as $r^{-2}$, and that $Q_{10}$ and
$a$ do not change as the dust moves outwards, we find that the optical depth found
via Eq.~3 scales as ${\rm R}_{in}^{-1}$. 

For what attains the tip of the AGB or the different post-AGB phases taken into account, 
we consider that both solid carbon dust and silicates form at a distance of $\sim 10~{\rm R}_*$
\citep{flavia12}, thus $\tau_{10}^{\rm onset} \sim (10~{\rm R}_*)^{-1}$. Regarding the dust presently 
surrounding the star, located at a distance ${\rm R}_{\rm in}$, we have 
$\tau_{10}^{\rm now} \sim {\rm R}_{\rm in}^{-1}$. Therefore, the relation we are looking for is

$$
\tau_{10}^{\rm now} \sim \tau_{10}^{\rm onset}\times (10~{\rm R}_*/{\rm R}_{\rm in})
\eqno(4)
$$

The comparison between the values of the optical depth derived via Eq.~4 and those
found in section \ref{sedfit} will allow testing the different hypothesis considered.
The methodology followed to understand how dust formation and mass loss work
during the final AGB phases and during the beginning of the post-AGB contraction will be also
based on the analysis of the derived expansion velocities with which the dust layer would travel 
from the evolutionary stage when it was released until now, in turn calculated based on the 
derived distance of the dusty region
and the time elapsed, found via the modelling of the AGB and post-AGB phases. 
Our reference point is that the wind velocity during the post-AGB phase is 
generally assumed to be comparable with the velocity in the AGB phase,
with values that typically span the $10-30$ km$/$s range \citep{he14, klochkova15}.

Whenever the results from this analysis prove not satisfactory, we will model the
post-AGB evolution by adopting different mass-loss rates, in the attempt of
understanding when the dust responsible for the observed infrared excess of the
various sources was released, and to derive the post-AGB mass loss that would
determine evolutionary time-scales consistent with the dynamical conditions
expected.

\section{Dust formation and wind propagation across the AGB-post-AGB frontier}
\label{wind}
A summary of the results derived in section \ref{sedfit} from the interpretation 
of the optical and infrared data of the stars considered are presented in Fig.~\ref{fall}, 
which shows $\tau_{10}$ of the individual sources as a function of 
the luminosity (left panel) and of the distance of the inner border of the dust layer 
from the centre of the star (right). Different trends can be detected in the figure, 
that correlate $\tau_{10}$ with the stellar luminosity and dust mineralogy, for stars 
of different metallicity.

For what concerns metal-poor carbon stars (full, red squares in the figure), 
we note: a) the correlation between luminosity and 
optical depth (left panel), the brighter stars exhibiting larger IR excesses;
b) luminosity and ${\rm R}_{\rm in}$ are anti-correlated, as the inner border of
the dusty layer is closer to the surface the brighter the star.

The metallicity of the stars is also playing a role in the present context:
the two higher metallicity carbon stars, indicated with open squares in Fig.~\ref{fall}, 
are characterized by optical depths significantly smaller than the lower metallicity 
counterparts with similar luminosities.

The oxygen rich stars behave differently from carbon stars.
They are all characterized by similar luminosities (see Fig.~\ref{ftracks}),
and optical depths extending over a factor $\sim 2$ with respect to the
average value $\tau_{10}=0.02$. Inspection of the right panel of Fig.~\ref{fall} shows 
that the dusty zone for these sources is closer than found for carbon stars.

To explain these trends, and to have a more exhaustive view of dust 
production and propagation during the phases from the tip of the AGB
until presently, we apply the methodology described in the previous section,
to study in detail a few objects, selected among those discussed in
section \ref{sedfit}, diverse in luminosity and dust composition.
The sources selected are J005803.08, J053250.69 and J050221.17. The first object,
discussed in section \ref{bright}, is a bright carbon star, which experienced several TDU 
episodes, responsible for the significant carbon enrichment in the surface regions. 
J053250.69, discussed in section \ref{lowmc}, is 
representative of faint, low-mass carbon stars, that reached the C-star stage during 
the last thermal pulses. J050221.17 is a low-mass, oxygen-rich star, which
failed to become carbon star,
because the number of TDU events experienced (if any) was not sufficient to reach the C$/$O$>1$ 
condition. We also discuss two carbon stars with sub-solar chemical
composition, and J052220.98, the low-luminosity carbon star, for which a possible
characterization was presented in section \ref{faintc}.

\subsection{Bright post-AGBs with carbon dust}
According to the discussion in section \ref{bright}, J005803.08 is
the brightest source investigated, with luminosity 
$\sim 12000~{\rm L}_{\odot}$, that indicates a $\sim 2.5~{\rm M}_{\odot}$
progenitor. This is close to the threshold limit required to start HBB \citep{ventura13}, 
which destroys the surface carbon and prevents the C-star stage. As the luminosity of
the star increases with the initial mass (see e.g. Fig.~\ref{ftracks}), we argue that J005803.08 is 
among the brightest post-AGBs with a carbon star chemistry. 

The AGB evolution of these stars was studied in details by \citet{flavia15a} and
more recently by \citet{ester21}, who outlined the large quantities of carbon accumulated 
in the surface regions and the intense dust production taking place, particularly 
during the very final AGB phases. Results from stellar evolution and dust formation
modelling suggest that at the tip of AGB this class of stars 
experiences mass loss rates slightly in excess of $10^{-4}~{\rm M}_{\odot}/$yr and efficient
dust production, with rates of the order of a few 
$10^{-6}~{\rm M}_{\odot}/$yr; a thick dust shell composed mainly of solid carbon is
expected to surround the star while evolving at the TAGB, with optical depths $\tau_{10} \sim 5$
\citep{flavia21}. At solar or 
sub-solar chemistries SiC grains might offer a significant contribution to the 
overall extinction of the radiation, but this is not the case for J005803.08, 
owing to the low metallicity.

\begin{figure}
\resizebox{1.\hsize}{!}{\includegraphics{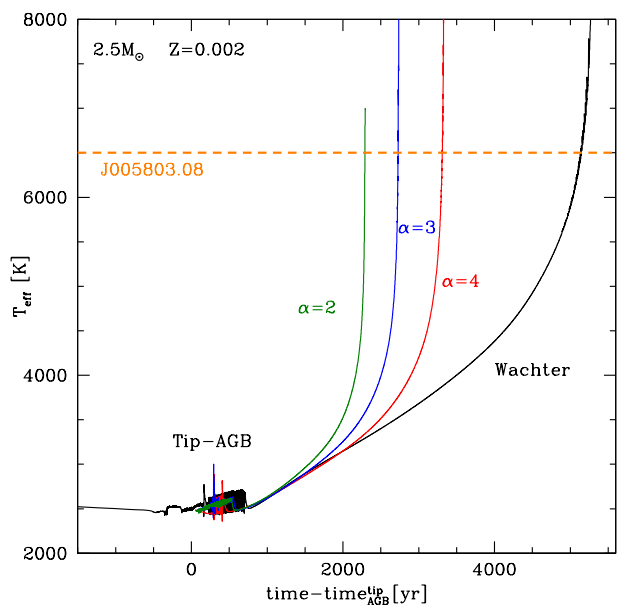}}
\vskip-60pt
\caption{Time variation of the effective temperature of a $2.5~{\rm M}_{\odot}$
model star during the transition from the AGB to the post-AGB phase. Times are
counted since the beginning of contraction. Black line indicates the standard
result, obtained by adopting the prescription from \citet{wachter02, wachter08}
for mass loss. Coloured tracks refer to results obtained by assuming
that $\dot M$ decreases as ${\rm T}_{\rm eff}^{-\alpha}$, with $\alpha=2$
(green), 3 (blue) and 4 (red). The orange, dashed line indicates the effective temperature 
of J005803.08 given in K14 and K15.} 
\label{f25ttef}
\end{figure}

\begin{figure}
\resizebox{1.\hsize}{!}{\includegraphics{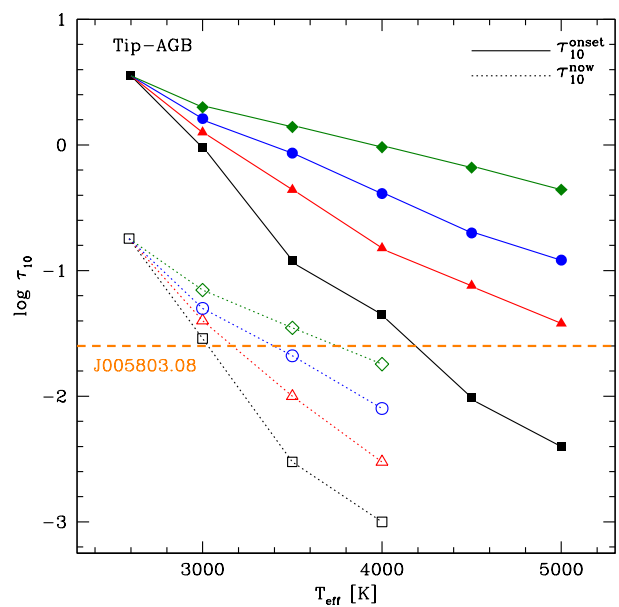}}
\vskip-60pt
\caption{Solid lines and full points indicate the values of the optical depth 
$\tau_{10}^{\rm onset}$ that characterize dust production at the tip of the AGB and 
during different phases along the contraction sequence of a $2.5~{\rm M}_{\odot}$ model star
(see text for details). The colors used correspond to different treatments of 
mass-loss, following the same colour coding as in Fig. \ref{f25ttef}. Dotted
lines and open symbols indicate the values of the optical depth $\tau_{10}^{\rm now}$,
that would characterize the dust responsible for the infrared excess observed nowadays,
at the effective temperature of J005803.08 (6500 K), under the assumption that it was 
released while the star was evolving through 
the evolutionary stages with the effective temperatures reported on the abscissa.
The dashed, orange line represents the optical depth obtained for J005803.08 from SED fitting.} 
\label{f2tau}
\end{figure}

Fig.~\ref{f25ttef} shows the time evolution of the effective temperature of
a $2.5~{\rm M}_{\odot}$ model star of metallicity similar to J005803.08, starting 
from the latest AGB phases through the contraction to the post-AGB. The
standard evolution, represented by the black line in the figure, was obtained 
with the input physics specified in section \ref{input} for the AGB 
and the post-AGB phases. In this case the time elapsed from the tip of the AGB until 
the current evolutionary stage, at $T_{\rm eff} = 6500$ K (see Tab.1), is $\sim 5600$ yr. 

For the tip of the AGB and for 5 further evolutionary phases selected 
during the post-AGB, corresponding to effective temperatures in the 
$3000~{\rm K}< {\rm T}_{\rm eff} < 5000~{\rm K}$ range, equally spaced
by $500$ K, we calculated the optical depth $\tau_{10}^{\rm onset}$. The
results are shown in Fig.~\ref{f2tau}. $\tau_{10}^{\rm onset}$ decreases as the star 
contracts (and the effective temperature increases), since 
the mass loss rate diminishes, thus the wind is less dense (see Eq.~2), so 
lower amounts of gas molecules are available to condense into dust. 

We followed the approach described in the previous section to calculate the corresponding 
$\tau_{10}^{\rm now}$ for each of the phases considered. We limited this analysis to the phases 
with ${\rm T}_{\rm eff} \leq 4000$ K, because dust production during later phases is too small 
to account for the IR excess observed (see Fig.~\ref{f2tau}).
To calculate $\tau_{10}^{\rm now}$ via Eq.~4 we used the 
distance of the inner border of the dusty zone derived in section \ref{bright} for J005803.08, 
i.e. ${\rm R}_{\rm in}=1.2\times 10^5~{\rm R}_{\odot}$, and the radius and optical
depth corresponding to each evolutionary phase. The results, indicated with black, open
squares in Fig.~\ref{f2tau}, are connected with a dotted line.

We are predisposed to rule out the possibility that the dust observed in the surroundings
of J005803.08 was released while the star was evolving at the tip of the AGB, for the
following reasons: a) the expected $\tau_{10}$ is a factor $\sim 7$ higher than that derived from SED fitting;
b) if we estimate the average expansion velocity
with which the dust would have travelled away from the star, on the basis of ${\rm R}_{\rm in}$ 
and of the time elapsed since the tip of the AGB to the current evolutionary status, 
we obtain $v \sim 0.5$ km$/$s, which is much lower than 
observed \citep{he14, klochkova15}.

In the context of the results obtained with the input given in section \ref{input}
(black squares in Fig.~\ref{f2tau}), the only possibility consistent with the 
analysis done in section \ref{sedfit} is that the dust observed now was released 
after the beginning of the contraction to the post-AGB, when the effective temperature of 
the star is $\sim 3000$ K. On the other hand we can rule out this possibility 
too, because the time elapsed since the evolutionary phase considered is 
$\sim 3600$ yr, which, combined with the distance of the dusty layer, leads to an average 
expansion velocity below 1 km$/$s, still too small. As a further motivation to disregard this possibility, 
we consider that if the winds and the dust move with a $\sim 0.8$ km$/$s velocity, the dust 
produced at the tip of the AGB would now be at a distance $\sim 2 {\rm R}_{\rm in}$, 
and would determine the formation of a bump in the SED of the star in the
$25-30~\mu$m region, which is not observed.

As we could not obtain any satisfactory overall explanation of the observations 
of J005803.08, we explored different mass-loss rate laws,
by changing the slope of $\dot{\rm M}$ with the effective temperature with respect to
the treatment by \citet{wachter02, wachter08}, according to which 
$\dot{\rm M} \sim {\rm T}_{\rm eff}^{-7}$. We calculated new evolutionary sequences
for the $2.5~{\rm M}_{\odot}$ model star, starting from the tip of AGB, where
we left the mass loss rate $\dot{\rm M}^{\rm TAGB}$ unchanged, then extended to the
post AGB, during which we assumed 
$\dot{\rm M} = \dot{\rm M}^{\rm TAGB} \times ({\rm T}_{\rm eff}/{\rm T}_{\rm eff}^{\rm TAGB})^{-\alpha}$, 
with $\alpha=2, 3, 4$. The results of these runs are reported 
in Fig.~\ref{f25ttef}, that shows the time variation of the effective temperature, and 
in Fig.~\ref{f2tau}, reporting the corresponding optical depths, with different colour coding. 
The evolutionary time scales get shorter the higher the mass loss rate, as the rate of the 
contraction process during the AGB to post-AGB transition is determined by the rate with which the 
residual envelope is lost \citep{iben83a}. The impact of the treatment of mass loss
on the time scales of the AGB to post-AGB transition was studied by \citet{vanhoof97},
who showed that the rate at which the effective temperatures of post-AGB stars
increase gets significantly shorter when it is assumed that the mass-loss rate keeps high
after the contraction phase begins.

Consistency between the expected and
the observed infrared excess (thus optical depth) is found in the
$\alpha=2$ and $\alpha=3$ cases, if we assume that the dust observed now was released
when the star contracted until the evolutionary phase characterized by effective temperatures
in the $3500-4000$ K range. In these cases the derived average expansion velocity of the
wind is in the $5-20$ km$/$s range, thus consistent with the observational evidence 
\citep{he14, klochkova15}. Furthermore, these velocities are sufficiently
fast that the dust released at the tip of AGB would in the meantime reach regions 
$5-20~{\rm R_{\rm in}}$ away from the central star, thus causing no effects on the 
SED in the $\lambda < 50~\mu$m spectral region.

In summary, the interpretation of the observations of J005803.08 indicate that
mass loss rates of bright carbon stars evolving at the tip of the AGB slightly in 
excess of $10^{-4}~{\rm M}_{\odot}/$yr are consistent with the infrared excess observed, 
provided that we assume that the dust was released when the effective temperatures
reached $\sim 3500-4000$ K, and that the rate of mass loss during the transition to
the post-AGB scales as ${\rm T}_{\rm eff}^{-2}$.

\begin{figure}
\resizebox{1.\hsize}{!}{\includegraphics{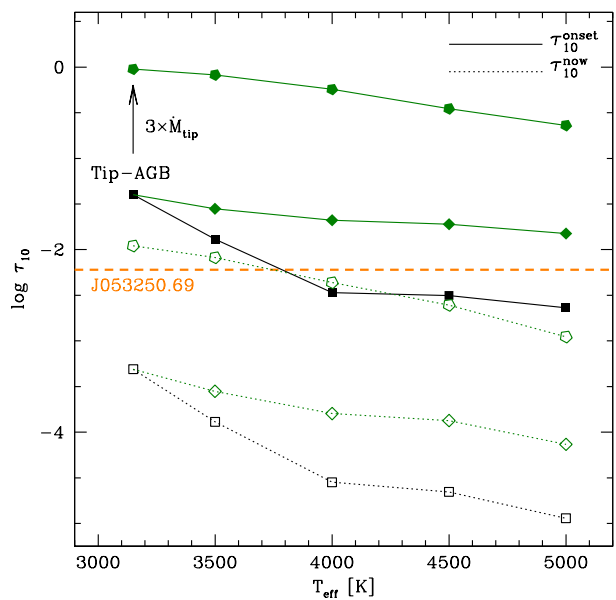}}
\vskip-60pt
\caption{Variation with the effective temperature of the optical
depth characterizing various evolutionary phases of a
$1~{\rm M}_{\odot}$ model star, from the tip of the AGB (first point 
on the plane) to the post-AGB. Full squares connected 
with solid lines indicate the values of $\tau_{10}^{\rm onset}$
obtained with the physical ingredients discussed in section 
\ref{input} to model the evolution of the star, whereas 
open squares refer to the expected optical depths that would
be observed in the present epoch, $\tau_{10}^{\rm now}$. Full and open diamonds refer
to the results obtained by assuming a different scaling of mass
loss with the effective temperature during the contraction to the
post-AGB (see text for details). Green pentagons indicate the optical 
depths obtained by artificially increasing the mass loss rate at the 
tip of AGB by a factor 3. The orange, dashed line indicates the optical
depth of J053250.69, obtained by SED fitting.} 
\label{f10tau}
\end{figure}

\subsection{Low-mass carbon stars}
J053250.69 is the faintest among the stars that we interpret
as surrounded by carbon dust, shown in Fig.~\ref{ftracks}, with the sole
exception of J052220.98, which will be discussed in a separate section.
In section \ref{lowmc} we interpreted this object as descending
from a low mass progenitor, with mass at the beginning of the
AGB phase of the order of $1~{\rm M}_{\odot}$. This mass is close to
the lower threshold required to reach the C-star stage.
As shown in Fig.~\ref{fg1}, which reports the time variation of
the surface C$/$O and $[$C$/$Fe$]$ of a $1~{\rm M}_{\odot}$ model
star with the same metallicity of J053250.69, the C-star stage is
reached during the last 2 TPs. The expected mass loss rate at the tip of the AGB
is $\dot {\rm M}^{\rm TAGB} \sim 10^{-5}~{\rm M}_{\odot}/$yr (see Fig.~\ref{fmlo}). 

By studying this source we are therefore exploring the low
luminosity tail of the carbon star population, in such a way that
the study of this star, coupled with the analysis of J005803.08
done previously, allows a thorough view of the late AGB and post-AGB
evolution of carbon stars. 

In analogy with the analysis of J005803.08, we considered the tip of the
AGB and different phases during the transition to the post-AGB of a
$\sim 1~{\rm M}_{\odot}$ model star, characterized by specific values of 
the effective temperature, selected in the $3500-5000$ K range, spaced by 500 K. 
We used the same methods 
described earlier to calculate $\tau_{10}^{\rm onset}$ for these stages. This analysis was applied 
both to the standard evolution of the $1~{\rm M}_{\odot}$ model star used to 
study J053250.69, with the mass loss description by \citet{wachter02, wachter08},
and to the case where the post-AGB mass loss rate is assumed to change as 
$\dot{\rm M} = \dot{\rm M}^{\rm TAGB} \times ({\rm T}_{\rm eff}/{\rm T}_{\rm eff}^{\rm TAGB})^{-2}$.

The results of this exploration are reported in Fig.~\ref{f10tau}, 
that shows with full symbols the values assumed by the optical
depth characterizing the SED of the star as it evolves from the tip
of the AGB (first point on the left side of the figure) to
the post-AGB. The corresponding
values assumed by the optical depth in the present epoch, 
under the hypothesis that the dust was released during the
evolutionary phases considered, are indicated with open symbols.
The results obtained by standard evolution modelling and those
based on the $\dot{\rm M} \sim {\rm T}_{\rm eff}^{-2}$ law
are indicated in Fig.~\ref{f10tau} with black squares and green diamonds, 
respectively.

We see in Fig.~\ref{f10tau} that the 
optical depths obtained are significantly smaller than those
derived for J053250.69, the differences 
exceeding a factor 10 in all cases. This conclusion holds independently of the
treatment of mass loss adopted to describe the contraction to the
post-AGB. 

A further issue is related to the estimated velocities with which the dusty
layer moved away from the star since it was released, until presently.
According to the standard description of the transition from the tip of the 
AGB to the post-AGB, the expected time transcribed from the tip of the AGB 
to the current phase would be $\sim 15000$ yr; this result, combined with 
the distance of the dusty region derived in section \ref{sedfit}, i.e. 
$2\times 10^5~{\rm R}_{\odot}$, corresponds to an average expansion velocity below 
1 km$/$s. Even if we assume that the dust was released during a later 
evolutionary phase, say when the effective temperature was $4000$ K
(that is however highly inconsistent with the estimated optical depth), the 
average expansion velocity would not exceed 1 km$/$s.

The conclusion drawn from this preliminary analysis is that the
mass loss rate at the tip of the AGB of low mass carbon stars
is underestimated. A higher
$\dot{\rm M}^{\rm TAGB}$ is required not only to obtain higher
optical depths, compatible with the infrared excess observed,
but also to shorten the contraction time scales, so that the derived
velocities are consistent with the observations \citep{he14, klochkova15}. 

Based on these arguments we calculated additional evolutionary sequences 
starting from the tip of the AGB, through the contraction to the post-AGB,
until reaching the effective temperature of J053250.69.
In these computations $\dot{\rm M}^{\rm TAGB}$ was artificially
enhanced with respect to the $\sim 10^{-5}~{\rm M}_{\odot}/$yr value 
given above, and the mass loss rate during the following contraction 
phases was changed accordingly, either by assuming the recipe by
\citet{wachter02, wachter08}, or by adopting the
$\dot{\rm M} \sim {\rm T}_{\rm eff}^{-2}$ relationship.

The results obtained by adopting a factor 3 increase in 
$\dot{\rm M}^{\rm TAGB}$ and $\dot{\rm M} \sim {\rm T}_{\rm eff}^{-2}$
for the following phases are indicated with green pentagons in 
Fig.~\ref{f10tau}. Consistency between the derived and the expected 
optical depth is obtained when assuming that the dust responsible for 
the currently observed infrared excess was released when the effective 
temperature of the star was $3500-4000$ K. In this case the time 
interval is reduced to $500-1000$ yr, thus the average expansion velocity of the 
wind is between 10 and 15 km$/$s.

We restrict the attention to the $\alpha=-2$ case. Indeed if we use the 
scaling of mass loss with effective temperature described in section 
\ref{input}, the only possibility to obtain an infrared excess comparable 
to the observed one is that the dust was released while the star was 
at the tip of the AGB. However, in that case the dust would have travelled 
for $\sim 5000$ yr, thus the average expansion velocity would be of the order of 1 
km$/$s.

\begin{figure}
\resizebox{1.\hsize}{!}{\includegraphics{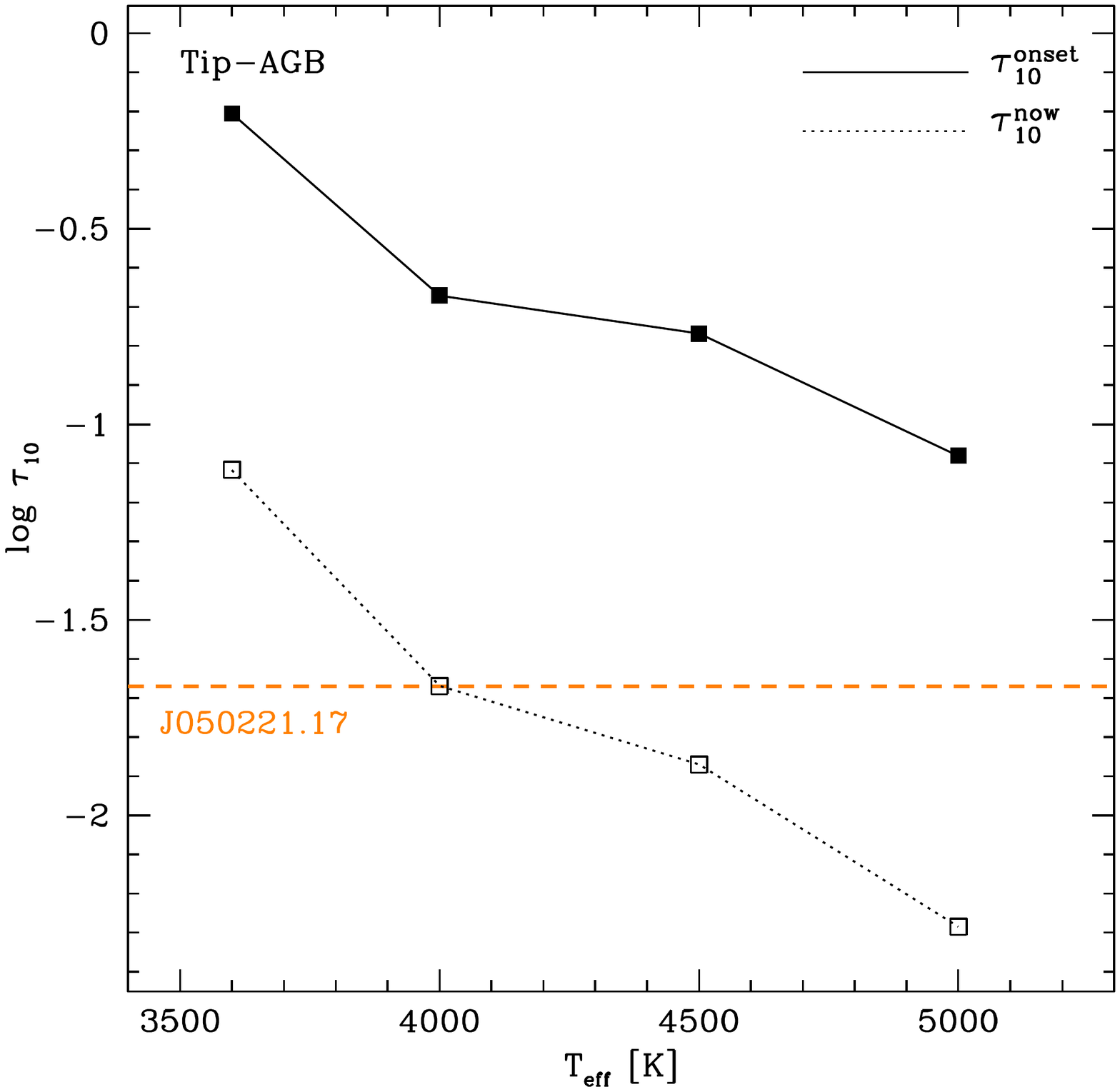}}
\vskip-60pt
\caption{Variation with the effective temperature of the optical
depth characterizing various evolutionary phases of a
$1~{\rm M}_{\odot}$ model star, from the tip of the AGB (first point 
on the plane) to the post-AGB. Full squares connected 
with solid lines indicate the values of $\tau_{10}^{\rm onset}$
obtained with the physical ingredients discussed in section 
\ref{input} to model the evolution of the star, whereas 
open squares refer to the expected optical depths that would
be observed in the present epoch, $\tau_{10}^{\rm now}$. The orange, dashed line indicates 
the optical depth of J050221.17, obtained by SED fitting.} 
\label{f08tau}
\end{figure}

\subsection{Carbon stars with sub-solar chemistry}
The carbon stars sample studied here encompass J050632.10 and J003643.94, 
two sub-solar metallicity stars with similar luminosities, around $6000~{\rm L}_{\odot}$
(see Tab.~\ref{tabero} and right panel of Fig.~\ref{ftracks}). 
In section \ref{z4m3} we suggested that these stars descend from 
$\sim 1-1.5~{\rm M}_{\odot}$ progenitors, which became carbon stars.

Both stars, indicated with open, red squares in Fig.~\ref{fall}, fall off the 
$\tau_{10}$ vs luminosity trend defined by low-metallicity carbon stars, as their optical
depths are $\sim 2$ times smaller; also, the current distances 
of the dusty region from the central star are the largest among the stars in the whole sample.

These differences can be at least partly explained by the fact that these two stars are
the hottest among the carbon stars considered, as reported in Tab.~\ref{tabero}. 
This indicates that they are evolving through a more advanced post-AGB phase in comparison with the
carbon stars of lower metallicity. This allowed the dusty layer to travel further away
and reach larger distances. 

However, the previous arguments alone are not sufficient to account for the low IR excess
observed in the SED of J050632.10 and J003643.94. Indeed, as shown in the right panel of 
Fig.~\ref{fall}, the distances of the dusty regions around these two sources are found to 
be $\sim 0.05-0.1$ dex higher than in the other carbon stars: by applying Eq.~4, we find that 
the corresponding differences in $\tau_{10}^{\rm now}$ are not expected to exceed $20-30\%$.
To recover the factor $\sim 2$ difference in the derived optical depth we must assume
that dust formation during the late evolutionary phases was significantly less efficient.

The most plausible explanation of the lower dust formation that J050632.10 and 
J003643.94 experienced during the late AGB phases is the smaller number of gaseous carbon molecules available 
to condense into solid particles in the surface regions, in turn related to the small
carbon excess with respet to oxygen. Indeed, the latter is the relevant quantity for the formation of 
solid carbon dust, given the high stability of CO molecules \citep{fg06}. The carbon excess
is intrinsically higher in metal poor stars $Z\sim 0.004$, given the lower initial oxygen content, while it
is smaller in stars with sub-solar chemistry, as the two sources investigated here. This quantity
is also sensitive to any possible oxygen enrichment connected to the TDU events experienced.

The latter point deserves particular attention. 
The results shown in Fig.~\ref{fg4} indicate that negligible oxygen enrichment is expected to 
have taken place in the surface regions of these two stars. On the other hand 
\citet{vanaarle13} claim significant oxygen enrichment for J050632.10, which implies a low 
carbon excess, and consequently poor dust production. We conclude that the low IR excess
characterizing the two stars considered is related to the metallicity and to the 
oxygen enrichment, witnessed by results from high-resolution spectroscopy.

There are some implications related
to these results, which will deserve further investigations in the future. First, the observations of 
J050632.10, similarly to other post-AGB stars in the Milky Way \citep{devika21}, outline a significant enrichment in
the surface oxygen, which is not expected based on the current modelling of the AGB phases. 
For what regards the description of mass loss of carbon stars,
the formalism must include some sensitivity to the surface chemical composition, as the efficiency
with which the winds of carbon stars are radiatively driven by dust is highly sensitive to the
carbon excess with respects to oxygen. Finally, in the description of carbon dust formation, it is 
recommended that the number density of seeds over which the dust grains grow, as suggested by
\citet{nanni13, nanni14}, is taken as
dependent on the carbon excess, instead of assuming a plain proportionality to the hydrogen
density, as done in most of the descriptions currently used.

\subsection{A faint, dusty C-star}
In the left panel of Fig.~\ref{fall} we note the anomalous position of
J052220.98, which falls off the general trend defined by C-stars:
despite the low luminosity, the smallest among the carbon stars in 
the sample, the optical depth derived in section \ref{sedfit} is 
similar to the brightest carbon stars considered.

In section \ref{faintc} we discussed the possible origin of this
object, and proposed that it experienced a late TP, which is the
reason for the relatively low luminosity. If this understanding 
proves correct, we could hardly apply the dust formation modelling
adopted in the previous cases so far discussed, given the poor
knowledge of the mass loss mechanism characterizing the tranition
to the post-AGB during and after the star experiences a late TP.
We speculate that the large infrared excess observed in the
SED of J052220.98 is caused by a fast removal of the residual envelope,
which favoured large mass loss rates, with enhanced dust production.

\subsection{Oxygen-rich, faint post-AGB stars}
\label{lowm}
We consider J050221.17 as representative of low-mass stars that
failed to reach the C-star stage. 
As discussed in section \ref{orich}, this source descends from
a low-mass progenitor, whose mass at the beginning of the AGB
phase was around $0.8~{\rm M}_{\odot}$. Results from stellar
evolution modelling indicate that such low-mass stars with 
the same metallicity as J050221.17 experience between 4 and 5
thermal pulses before their external envelope, initially of mass
$\sim 0.3~{\rm M}_{\odot}$, is lost, and the contraction to the
post-AGB phase starts. The surface chemistry is substantially
unchanged since the beginning of the AGB evolution.
Similarly to the
carbon stars discussed earlier, the largest rates of mass loss, which
in this particular case are of the order of $10^{-6}~{\rm M}_{\odot}/$yr,
are reached at the tip of the AGB. 

The dust formed in the wind of these stars is
mainly composed by silicates, with alumina dust
contributing for less than $10\%$ \citep{flavia14a}.
Following the method described in section \ref{input}, we find that the optical 
depth at the tip of AGB is $\tau_{10}=0.5$.

We modelled dust formation in the wind for different points
taken along the evolutionary track, following the same criterion,
based on the values of the effective temperature, chosen for carbon 
stars: in this specific case, considering that the tip effective temperature
is $\sim 3600$ K, we considered ${\rm T}_{\rm eff}=4000, 4500, 5000$ K.
In analogy with the study of carbon stars presented earlier in this 
section, we use the radii of the star during each of the phases
considered, in combination with the distance of the dust layer, derived
for J050221.17 in section \ref{sedfit}, namely 
${\rm R}_{\rm in} = 5.45\times 10^4~{\rm R}_{\odot}$, to calculate the current optical
depth, via Eq.~4. The results are shown in Fig.~\ref{f08tau}, where we report
the values of $\tau_{10}^{\rm now}$ as a function of the effective 
temperature.

The expected duration of the evolution of the star between the
tip of the AGB and nowadays is $\sim 60000$ yr. If we consider 
the current distance of the dusty region from the central star ${\rm R}_{\rm in}$,
we find that the dust particles
should move extremely slowly since they are released, with 
average velocities below 0.1 km$/$s. This is main reason why 
we believe extremely unlikely that the dust responsible for the
infrared excess currently observed was released while the star
was evolving at the tip of AGB. A further motivation to disregard
this possibility is that the derived $\tau_{10}^{\rm now}$ is
significantly higher than derived in section \ref{sedfit}.

Indeed the results shown in Fig.~\ref{f08tau} indicate that
consistency between the expected and the derived $\tau_{10}$
are found under the hypothesis that the dust was released
when the effective temperature was in the $4000-4500$ K
range. In the latter case we should assume that the time
elapsed since the dust was released is substantially shorter,
which would correspond to average velocities $\sim 0.5-1$ km$/$s.
These velocities are lower than found during the study of the
wind of carbon stars, which according to our understanding is the
reason why in the case of M-type post-AGBs the distance between the
dusty cloud and the central object is smaller than for C-rich
stars, as clear in the right panel of Fig.~\ref{fall}. 
The generally lower extinction coefficients of silicates with respect
to carbonaceous dust species is the main reason for the lower expansion velocity
with which the dust moved away from O-rich stars than for the C-rich
counterparts. Furthermore, M-type stars are the faintest
objects in the sample studied here, thus the effects of the radiation
pressure on the dust particles, which is proportional to the luminosity
of the star, are lower.

\subsection{An overall view of dust surrounding post-AGB stars}
The results presented in this section allow us to interpret the findings
reported in Fig.~\ref{fall}. The increasing trend of the optical depth
with luminosity (hence mass) traced by metal-poor carbon stars is 
mainly due to the fact that brighter stars evolve faster during the AGB-post-AGB
transition, thus the dusty region is closer to the stellar surface. The 
higher carbon dust production rates of the stars of higher mass is
also playing a role in this context. 

Oxygen-rich stars that failed to reach the
C-star stage cover a narrower range of luminosities than the C-rich counterparts,
thus no significant trends can be detected. The dust production rate during the
late evolutionary phases is significantly smaller than for carbon stars, thus
the acceleration experienced by the outflow is lower, and the current distances
of the dusty regions from the photosphere are shorter. 

The SEDs of carbon-rich post-AGBs
with sub-solar chemistry ($Z\sim0.004$) present smaller infrared excesses with respect to the
lower-metallicity counterparts, owing to the scarcity of gaseous carbon molecules
available to form dust, in turn related to the large surface oxygen abundances.

\section{Conclusions}
\label{concl}
We study a sample of single stars identified as post-AGB sources observed in the
Magellanic Clouds, with the goal of providing a characterization in terms of
mass and formation epoch of the progenitors, and of the mineralogy of the dust 
responsible for the 
observed infrared excess. The analysis is based on the
combination of optical and near infrared data, that allow to trace the
morphology of the spectral energy distribution.

The comparison between the observations and synthetic SED modelling obtained 
via the radiative transfer code DUSTY allowed to
identify a majority of stars surrounded by carbonaceous dust, and 
five objects with silicates in the surroundings. 
The determination of the luminosity, still obtained within the 
SED fitting process, led to the derivation of the initial mass of the
progenitors, based on the tight link between core mass and luminosity that
characterize post-AGB stars. Overall, we find that the sub-sample selected
is composed by stars descending from progenitors with mass in the
$0.7-2.5~{\rm M}_{\odot}$ range, formed in epochs ranging from
500 Myr to 8 Gyr ago. 

For some of the stars considered, the interpretation proposed here finds
a valid confirmation based on the results from high-resolution spectroscopy, which 
nicely agree with the expectations from stellar evolution modelling. The only
exception is a star with luminosity $\sim 4500~{\rm L}_{\odot}$, inconsistent with 
the presence of carbon dust in the surroundings, for which we propose an alternative
explanation, based on the occurrence of a late thermal pulse. Overall, the 
present analysis indicates that the results from the present generation of AGB models
are consistent with the post-AGB observations, as far as the general pattern of
the surface chemistry with luminosity is concerned. This agreement demands
further investigation before it can be confirmed, given the small number
of sources investigated.

The analysis of the infrared excess observed, combined with 
results from evolutionary modelling of the late AGB and post-AGB phases, 
proves a valuable tool to understand the physical behaviour of the stars
before and during the AGB - post-AGB transition, as well as to study the
dynamics of the outflow, moving away from the central stars dust currently
observed was released.

The general conclusion drawn in this context is that the dust nowadays
observed around post-AGB stars was released after contraction to post-AGB
began, and the effective temperatures increased to $\sim 4000$ K. 
The infrared excess observed is primarily determined by the time scale of
the post-AGB evolution, which proves crucial to determine the current
location of the dusty region, and the dust production rate experienced
when the stars evolve through the tip of the AGB and the early post-AGB.

In the case of carbon stars the infrared excess increases with the luminosity of
the stars, as bright stars evolve faster and are expected to experience
a more efficient formation of dust. Detailed comparison with stellar
evolution modelling indicates that the mass loss and dust formation rates
of $M \geq 2~{\rm M}_{\odot}$ stars are correctly described, whereas
in the low mass domain a factor $\sim 3$ increase is required. 
A further indication from the present study, connected with the small
infrared excess characterizing the SED of a few low-mass stars of sub-solar 
chemical composition, is the need of a mass loss prescription that is
sensitive to the surface chemistry, particularly to the carbon excess with
respect to oxygen.

The oxygen-rich stars in the sample are interpreted as the progeny of low-mass
stars, that failed to reach the C-star stage. The dust in their surroundings
is mostly composed by silicates in a crystalline form. In this case satisfactory
agreement is found with the theoretical expectations, which indicate mass loss
rates at the tip of the AGB of the order of $10^{-6}~{\rm M}_{\odot}/$yr.
The dusty shells of M-type stars are found to be on the average closer to
the central star with respect to carbon stars, which we interpret as a result
of the lower radiation pressure acting on silicates than on carbon dust,
which results in slower winds. 

As a next step in our research, we intend to extend our study to the Galactic sample of post-AGB stars presented in \citet{devika22a}. The upcoming release of the Gaia DR3 astrometric data will be instrumental in further capitalising on the Galactic sample of objects. Additionally, as future work, we will aim to obtain robust observational evidence to further constrain the dust chemical composition of our target sample. In this regard, high-resolution IR and mid-IR spectra from James Webb Space Telescope (JWST), covering the $\sim$\,5\,$-$\,28\,$\mu$m wavelength regime, will be particularly useful since the spectra of post-AGB stars show interesting features such as atomic fine-structure and hydrogen lines, PAHs, crystalline and amorphous silicates, SiC, MgS, the enigmatic 21$\mu$m feature, alumina, C$_{\rm 2}$H$_{\rm_2}$, SiS, SiO, TiO and H$_{\rm 2}$O \citep{sloan16,gielen11}. The intermingling of these observations with our newly developed models will allow us to further develop our description of dust formation in the outflows from post-AGB stars.

\label{end}

\begin{acknowledgements}
DK  acknowledges  the  support  of  the  Australian  Research Council (ARC)  Discovery  Early  Career  Research  Award (DECRA) grant (DE190100813). This research was supported in part by the Australian Research Council Centre of Excellence for All Sky Astrophysics in 3 Dimensions (ASTRO 3D), through project number CE170100013. HVW acknowledges support from the Research Council of the KU Leuven under grant number C14/17/082. EM acknowledges support from the INAF research project “LBT - Supporto Arizona Italia".

\end{acknowledgements}

%
%

\end{document}